\documentclass[a4paper,fleqn,usenatbib]{mnras}

\usepackage{newtxtext,newtxmath}
\usepackage[T1]{fontenc}
\usepackage{ae,aecompl}

\usepackage{graphicx}	% Including figure files
\usepackage{amsmath}	% Advanced maths commands
\usepackage{amssymb}	% Extra maths symbols
\usepackage{xcolor}
\usepackage{enumitem}
\usepackage{subfig}
\usepackage[para]{threeparttable}
\usepackage{footmisc}
\usepackage{times}
\title[Dust properties in NGC\,628]{Revealing the dust attenuation properties on resolved scales in NGC\,628 with SWIFT UVOT data}

\author[M. Decleir et al.]{Marjorie Decleir$^{1}$\thanks{E-mail: marjorie.decleir@ugent.be},
Ilse De Looze$^{1,2}$,
M\'ed\'eric Boquien$^{3}$,
Maarten Baes$^{1}$,
\newauthor Sam Verstocken$^{1}$,
Daniela Calzetti$^{4}$,
Laure Ciesla$^{5}$,
Jacopo Fritz$^{6}$,
Rob Kennicutt$^{7}$,
\newauthor Angelos Nersesian$^{1,8,9}$,
Mathew Page$^{10}$
\\
$^{1}$Sterrenkundig Observatorium, Universiteit Gent, Krijgslaan 281 S9, 9000 Gent, Belgium\\
$^{2}$Department of Physics \& Astronomy, University College London, Gower Street, London WC1E 6BT, UK\\
$^{3}$Universidad de Antofagasta, Centro de Astronom\'ia, Avenida Angamos 601, Antofagasta 1270300, Chile\\
$^{4}$Department of Astronomy, University of Massachusetts -- Amherst, Amherst, MA 01003, USA\\
$^{5}$Laboratoire AIM-Paris-Saclay, CEA/DSM/Irfu - CNRS - Universit\'e Paris Diderot, CEA-Saclay, F-91191 Gif-sur-Yvette, France\\
$^{6}$Instituto de Radioastronom\'ia y Astrof\'isica, UNAM, Campus Morelia, AP 3-72, 58089, Michoac\'an, Mexico\\
$^{7}$Steward Observatory, University of Arizona, 933 N Cherry Avenue, Tucson, AZ  85721-0065, United States\\
$^{8}$Department of Astrophysics, Astronomy \& Mechanics, Faculty of Physics, University of Athens, 
Panepistimiopolis, GR15784 Zografos, Athens, Greece\\
$^{9}$Institute of Astronomy, Astrophysics, Space Applications \& Remote Sensing, National Observatory of Athens, GR15236 Penteli, Greece\\
$^{10}$Mullard Space Science Laboratory, University College London, Holmbury St Mary, Dorking, Surrey RH5 6NT, UK}

\date{Accepted XXX. Received YYY; in original form ZZZ}

\pubyear{2018}

\begin{document}
\label{firstpage}
\pagerange{\pageref{firstpage}--\pageref{lastpage}}
\maketitle
\textbf{}
\begin{abstract}
Understanding how dust attenuation laws vary between and within galaxies is a key question if we want to reliably measure the physical properties of galaxies at both global and local scales. To shed new light on this question, we present a detailed study of the slope and bump strength of the attenuation law in the nearby spiral galaxy NGC\,628 at the resolved spatial scale of 325\,pc. To do so, we have modelled a broad multi-wavelength dataset from the ultraviolet (UV) to the infrared (IR) with the state-of-the-art SED fitting code \textsc{cigale}, including SWIFT UVOT data for which we have developed a new optimized reduction pipeline. We find that the median dust attenuation curve of NGC\,628 is fairly steep, but not as steep as the SMC curve, and has a sub-MW-type UV bump. We observe intriguing variations within the galaxy, with regions of high $A_{\text{V}}$ exhibiting a shallower attenuation curve. We argue that the flattening of the curve is due to a dominance of absorption-over-scattering events at higher $A_{\text{V}}$. No trend between the bump strength and the IRAC\:8.0\,\micron\ emission was found. However, this does not necessarily rule out PAHs as the main contributors to the UV bump.
\end{abstract}


%%%%%%%%%%%%%%%%%%%%%%%%%%%%%%%%%%%%%%%%%%%%%%%%%%

%%%%%%%%%%%%%%%%% BODY OF PAPER %%%%%%%%%%%%%%%%%%

\section{Introduction}
Interstellar dust only makes up a very small fraction of the interstellar mass in galaxies (typically about 1\% in metal-rich spirals or less in lower-metallicity galaxies, e.g. \citealt{2014A&A...563A..31R}). Nevertheless, its impact on the other baryonic components in galaxies should not be underestimated, since dust regulates several physical and chemical processes. For example, dust particles act as a catalyst for the formation of molecular hydrogen \citep{1963ApJ...138..393G}, regulate the heating of the neutral gas component through photoelectric heating \citep{1978ApJS...36..595D} and inelastic interactions, and provide shielding for molecules from the hard ultraviolet (UV) radiation of young stars \citep{2017MNRAS.467..699H}. Furthermore, dust is a reservoir of metals, which are the main building blocks for planet formation, and in fact for all organic material that we know today. Finally, as dust grains absorb and scatter about 30\% of the starlight (UV and optical radiation) in the Universe \citep{2005ARA&A..43..727L,2016A&A...586A..13V}, they heavily influence our view on the other galaxy components. To obtain an undistorted view of galaxies, understanding the interplay between dust and starlight is essential.

It is important at this point to stress the difference between extinction and attenuation. Extinction is caused by dust absorption and scattering out of the line-of-sight. Extinction curves describe these effects of dust on the galaxy's luminosity at different wavelengths for the specific geometry of a single star behind a screen of dust.  For all other geometries the effect of dust can be described by a dust attenuation curve, which also includes scattering into the line-of-sight due to the mix of stars and dust.

\begin{figure}
	\includegraphics[width=\columnwidth]{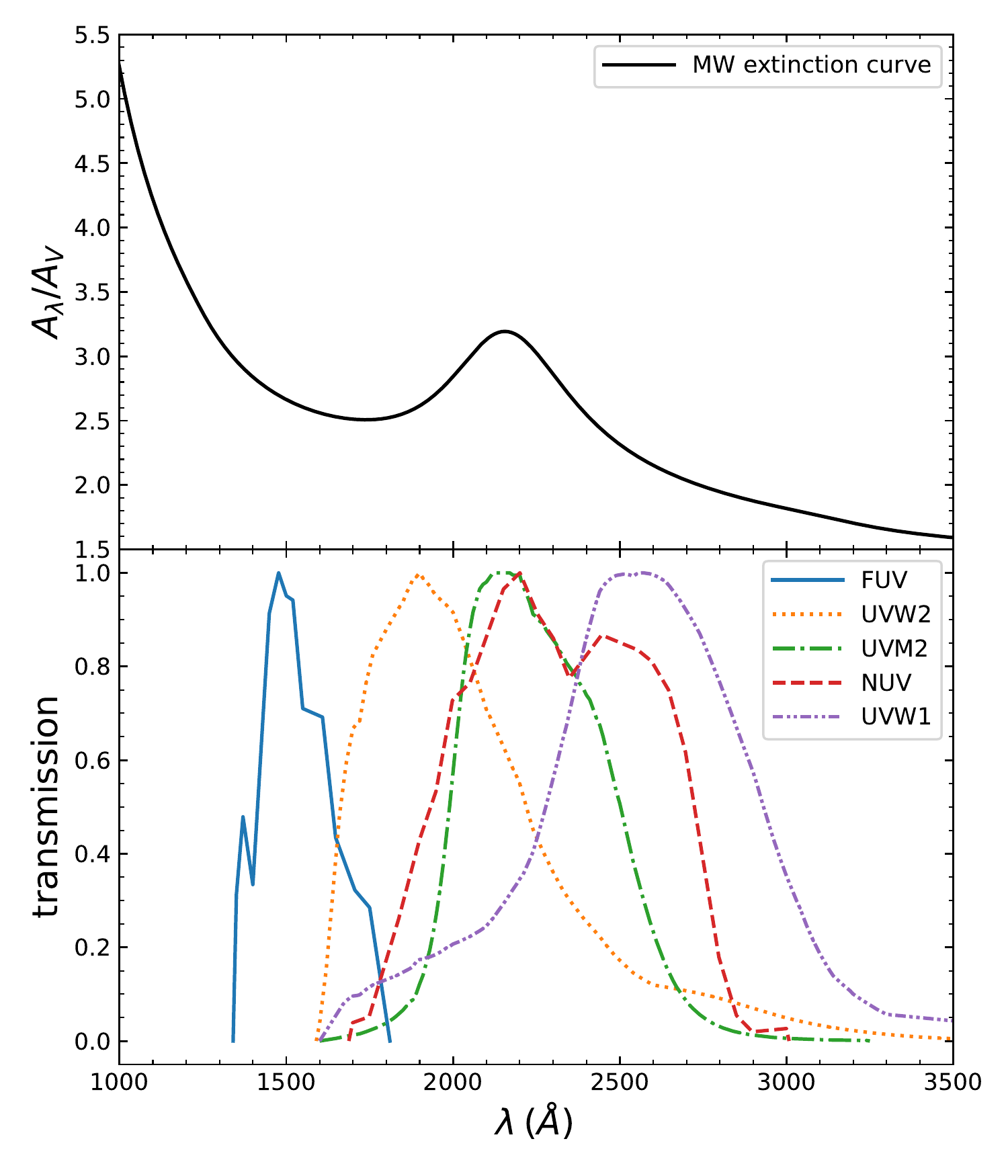}
    \caption{Transmission curves of the SWIFT UV filters and the GALEX filters, taken from the SVO Filter Profile Service (\url{http://svo2.cab.inta-csic.es/svo/theory/fps/}). The MW extinction curve in the top panel is taken from \citet{1989ApJ...345..245C}.}
    \label{fig: uvot_filters}
\end{figure}

The specific shape of a dust extinction (or attenuation) law is characterized by the steepness of its slope in UV, optical and (near-)infrared (NIR) wavebands, and the strength of the 2175\,\AA\ bump (i.e. an excess in dust absorption around that wavelength). Often, the slope is parametrized through the total-to-selective extinction ratio, $R_{\text{V}}=A_{\text{V}}/E(B-V)$, with an average value of $R_{\text{V}}=3.1$ (in the Milky Way, MW, \citealt{1989ApJ...345..245C}), and with regions dominated by smaller/larger grains characterized by steeper/shallower slopes (or thus smaller/larger $R_{\text{V}}$). The steepness of the average Small Magellanic Cloud dust law ($R_{\text{V}}=2.7$, \citealt{2003ApJ...594..279G}) is therefore thought to result from a grain population with predominantly smaller grains \citep{2003ApJ...588..871C,2005ApJ...630..355C}. The bump feature is believed to be caused by small grains with a carbonaceous composition (e.g. polycyclic aromatic hydrocarbons (PAHs) or graphite grains), based on laboratory studies, although an observational link between the abundance of possible bump carriers and bump strength remains to be found (e.g. \citealt{2017MNRAS.466.4540H}). Also, the invariable peak wavelength but changing width of the bump feature (e.g. \citealt{2004ApJ...616..912V}) has been proven hard to reconcile with theoretical calculations \citep{1993ApJ...414..632D}. A strong 2175\,\AA\ bump is observed in the MW (see Fig. \ref{fig: uvot_filters}), with considerable variations from one sightline to the other, while only a weak bump is present in the Large Magellanic Cloud (LMC) and it is completely absent in the Small Magellanic Cloud (SMC). Interestingly, dust attenuation laws in starburst galaxies also tend to be characterized by a weak bump or no bump at all \citep{2000ApJ...533..682C}. However, in addition to varying grain properties, different geometrical distributions are argued to be able to affect the presence of a bump feature (e.g. \citealt{2007MNRAS.375..640P}).

In Local Group galaxies, dust extinction curves can be inferred from a comparison of two stars of the same spectral type but affected by different levels of extinction (this technique is referred to as the ``pair method''). The diversity of dust extinction laws derived through the pair method along different sightlines \citep[e.g.][]{2003ApJ...594..279G,2015ApJ...815...14C} is consistent with a non-uniform population of interstellar dust grains in a single galaxy. For galaxies beyond the Local Group, on the other hand, (with the exception of pairs of occulting galaxies, e.g. \citealt{2013MNRAS.433...47H,2014AJ....147...44K}), we resort to deriving dust attenuation curves, which hold information on both the dust absorption and scattering out of the line-of-sight (characterized by optical grain properties), and the process of scattering into the line-of-sight (dependent on the relative star-dust geometry). As nicely summarized by \cite{2018ApJ...859...11S}, dust attenuation curves are typically inferred using two distinct methods; either through a statistical ``empirical'' comparison of large samples of galaxies with different levels of extinction (assuming that the average dust attenuation law does not vary significantly among galaxies), or alternatively through a ``model'' comparison of the observed stellar radiation with models of stellar populations attenuated by a range of dust attenuation laws. Due to the computational cost of this last ``model'' method, the first dust attenuation curves have been inferred based on the ``empirical'' comparison method \cite[e.g.][]{1994ApJ...429..582C,2000ApJ...533..682C}. Similar techniques have also been used in more recent works \cite[e.g.][]{2011MNRAS.417.1760W,2016ApJ...818...13B,2017ApJ...840..109B}. Owing to the introduction of Bayesian inference approaches and other statistical methods, capable of doing a rigorous search of the full parameter space, several authors have published ``model'' inferred dust attenuation curves \cite[e.g.][]{2010ApJ...718..184C,2011A&A...533A..93B,2012A&A...545A.141B,2017ApJ...837..170L,2018ApJ...859...11S} in recent years. Intriguingly, the dust attenuation laws inferred from ``empirical'' and ``model'' methods do not always agree, and potentially result in very discrepant dust properties. This apparent dichotomy was attributed to the assumption of a uniform dust attenuation law across entire galaxy samples, and the underestimation of dust attenuation for galaxies considered to be completely transparent, in the case of the ``empirical'' comparison method \citep{2018ApJ...859...11S}.

Due to a lack of detailed knowledge about the dust properties outside our own Galaxy and the Magellanic Clouds, in extragalactic studies they are generally assumed to be the same as in the MW. Coupling these dust grain properties to different geometries results in different attenuation laws. For normal galaxies, the MW extinction law is usually adopted, while the \citet{2000ApJ...533..682C} relation is used for star-bursting galaxies. However, there is growing evidence for strong deviations from a universal dust attenuation law, based on the different shapes of dust attenuation curves observed in other galaxies \citep[e.g.][]{2018ApJ...859...11S,2017ApJ...840..109B,2017ApJ...851...90B,2015ApJ...806..259R,2016ApJ...827...20S}. Imprecise assumptions about the shape of the dust attenuation curve, which is dependent on both the dust properties and the relative star-dust geometry, can result in unattenuated stellar luminosities erroneous by a factor of a few \citep{2014A&A...561A..39B,2007MNRAS.379.1022D}. Dust attenuation is still one of the most uncertain parameters to recover intrinsic star formation rates (SFRs), star formation histories (SFHs), dust masses and stellar masses at all redshifts \citep{2014ARA&A..52..415M}.

Statistical studies of large galaxy samples allow to infer an average dust attenuation law, to study the spread of galaxies from this mean trend, and to understand how the dust attenuation law varies with galaxy properties. However, to understand what physical processes are driving dust evolution on local scales within galaxies, we require spatially resolved studies of the dust attenuation law in a set of nearby galaxies. To this aim, we launched the DustKING project which will study the attenuation curve in the KINGFISH (Key Insights on Nearby Galaxies: a Far-Infrared Survey with Herschel, \citealt{2011PASP..123.1347K}) sample of sixty-one well-resolved nearby galaxies, spanning a wide range of morphologies and star formation activities, and thus probing different stages in dust evolution. The KINGFISH sample largely benefits from an extensive ancillary set of imaging and spectroscopy data. In this work, we study the dust attenuation properties on spatially resolved scales of about 325\,pc (or 7$\arcsec$) in NGC\,628 (or M74), a nearby ($D \approx 9.59\,\textrm{Mpc}$, \citealt{2017ApJ...834..174K}) grand design spiral galaxy in the constellation Pisces. This galaxy has been classified as an SAc spiral galaxy with a stellar mass of $\textrm{log} (M_*/M_{\odot}) = 9.821$, a $\textrm{SFR} = 1.07 \:M_{\odot}/yr$ and a metallicity of $12+\text{log}(O/H)=8.80$ \citep{2018arXiv180904088H}. The main motivation to start with NGC\,628 is that it is a well-studied and well-resolved galaxy, for which a lot of ancillary data (including deep SWIFT images, see below) is available. Although we are studying the attenuation properties on resolved scales within the galaxy, it should be noted that there might still be large variations in attenuation within each pixel. Since molecular clouds in nearby galaxies are found to be between 4 and 190\,pc in radius \citep{2013ApJ...779...46H}, one resolved region (of about 325\,pc) can contain a mix of clouds with different gas and dust properties. This can affect the observed dust attenuation curve. 

In our study, we use a multi-wavelength dataset, ranging from the far-ultraviolet (FUV) to the far-infrared (FIR), as described in more detail in section \ref{sec: Data}. In particular, we make use of UV images taken by the SWIFT satellite. The SWIFT spacecraft \citep{2004ApJ...611.1005G} is designed to detect events such as gamma-ray bursts and is, therefore, equipped with three photon-counting instruments which are sensitive to single-photon events of point sources. The Burst Alert Telescope (BAT, 15-150 keV) and the X-ray Telescope (XRT, 0.3-10 keV) focus on detecting high-energy photons, while the UV/Optical Telescope (UVOT, 1700-6000\,\AA, \citealt{Roming2005}) probes the stellar emission in the UV/optical wavelength domain. The set of three UVOT near-ultraviolet (NUV) broadband filters (UVW2, UVM2 and UVW1 with effective wavelengths of 1991, 2221 and 2486\,\AA\ respectively) are uniquely suited to constrain the attenuation curve. Fig. \ref{fig: uvot_filters} shows the transmission curves of these filters and the GALEX FUV and NUV filters, together with an average MW extinction curve. More specifically, the UVM2 filter overlaps the 2175\,\AA\ dust absorption feature, so when the UVM2 flux is suppressed relative to that of UVW2 and UVW1, the degree of suppression traces the amplitude of the bump. Likewise, the amount of UVW2 flux that is extinguished compared to UVW1 helps to trace the slope of the curve, especially when combined with optical and NIR observations \citep{2017MNRAS.466.4540H}. Additionally, the images have a high spatial resolution that is twice as good as the resolution of GALEX. Since the SWIFT telescope is designed to detect point sources, the standard SWIFT data reduction pipeline is not suitable for our purposes (i.e. the study of the dust on resolved scales in extended sources such as nearby galaxies). In order to solve this issue, we developed a new reduction pipeline, which is entirely optimized for extended sources. We present our pipeline in section \ref{sec: Data reduction}.

The SWIFT data have, so far, only been used to study the interstellar dust properties in a handful of nearby galaxies: M81 and Holmberg IX \citep{2011AJ....141..205H}, both exhibiting grain characteristics that are not very different from the MW; the starburst galaxy M82 \citep{2014MNRAS.440..150H,2015MNRAS.452.1412H} which shows a gradient in the dust attenuation properties with projected galactocentric distance; the SMC \citep{2017MNRAS.466.4540H} which seems to have a large-scale gradient in bump strengths; and the central 200 pc of the M31 bulge \citep{2014ApJ...785..136D}, where the extinction curve is found to be steep ($R_{\text{V}}= 2.4-2.5$). Another work by De Looze et al. (in prep.) will do a study similar to ours for M51.

The remainder of this paper is structured as follows. Section \ref{sec: Data} describes the data that was used. Sections \ref{sec: Data reduction} and \ref{sec: Data processing} focus on the reduction pipeline of the SWIFT UVOT data and the further processing of the images respectively. In section \ref{sec: Colour plots} we discuss empirical UV colour-colour plots. Section \ref{sec: CIGALE} explains the modelling of the galaxy with \textsc{cigale} and presents our results, which are further discussed and compared to other studies in section \ref{sec: Discussion}. Finally, section \ref{sec: Summary} summarizes our study and main results. The appendix of this paper contains the following sections: appendix \ref{app: ID} with the IDs of the SWIFT UVOT images used in this work, appendix \ref{app: coicorr} with a detailed description on the calculation of the coincidence loss correction for the SWIFT UVOT images, appendix \ref{app: cigale_par} lists the parameter values used in the \textsc{cigale} fitting, appendix \ref{app: figures} shows the observations and models of the \textsc{cigale} fitting in each waveband and appendix \ref{app: mock} describes the mock data analysis.

\section{Data}
\label{sec: Data}
In this work, we employ multi-wavelength data ranging from the FUV to the FIR, as listed in Table \ref{tab:filters}. We are not using images with a longer wavelength (which have a poorer resolution) than the PACS\:100\,\micron\ image since we want to perform our study on the smallest possible resolved scales (of about 7\arcsec). As discussed in the introduction, we use images from the SWIFT UVOT instrument for which we developed a new data reduction pipeline optimized for extended sources. More details about the SWIFT data and the reduction pipeline are given in section \ref{sec: Data reduction}.

The MIPS\:24\,\micron\ image was obtained from SINGS (Spitzer Infrared Nearby Galaxies Survey) and the data reduction of this image is described in \cite{2003PASP..115..928K}. All other images were taken from the DustPedia Archive\footnote{\url{http://dustpedia.astro.noa.gr}}, which provides access to multi-wavelength imagery and photometry for 875 nearby galaxies as well as model derived physical parameters for each galaxy \citep{2017PASP..129d4102D}. The DustPedia images were reduced in a homogeneous way as described in detail in \citet{2018A&A...609A..37C}. Further processing of the data is discussed in section \ref{sec: Data processing}.

\begin{table*}
	\centering
	\caption{Information about the dataset used in our study. Column 1: Filter name. Column 2: Effective wavelength $\lambda_{\text{eff}}$. Column 3: Resolution (full width at half maximum, FWHM), taken from \citet{2010MNRAS.406.1687B} for the SWIFT bands and from \citet{2018A&A...609A..37C} for all other bands. Column 4: Galactic foreground extinction $A_{\lambda}/A_V$, obtained by convolving the MW extinction curve with the transmission curve of the filter. Column 5: Calibration uncertainty. Column 6: Reference for the calibration uncertainties. Column 7: Typical background noise.}
	\label{tab:filters}
    \begin{threeparttable}
	\begin{tabular}{lllllcl}
		\hline
		Filter & $\lambda_{\text{eff}}$ & FWHM (\arcsec) & $A_{\lambda}/A_V$ & cal. unc. (\%) & Ref. cal. & sky noise (\%)\\
		\hline
		GALEX FUV & 1528 \AA & 4.3 & $2.65 \pm 0.02$ & 4.5 & (a) & 1.2\\
		SWIFT UVW2 & 1991 \AA & 2.92 & $2.67 \pm 0.04$ & 2.8 & (b) & 2.6\\
		SWIFT UVM2 & 2221 \AA & 2.45 & $2.80 \pm 0.05$ & 2.8 & (b) & 2.5\\
		GALEX NUV & 2271 \AA & 5.3 & $2.63 \pm 0.07$ & 2.7 & (a) & 1.2\\
		SWIFT UVW1 & 2486 \AA & 2.37 & $2.26 \pm 0.13$ & 2.8 & (b) & 3.2\\
		SDSS u & 3551 \AA & 1.3 & $1.59 \pm 0.01$ & 1.3 & (c) & 8.4\\
		SDSS g & 4686 \AA & 1.3 & $1.19 \pm 0.02$ & 0.8 & (c) & 1.2\\
		SDSS r & 6165 \AA & 1.3 & $0.868 \pm 0.007$ & 0.8 & (c) & 1.2\\
		SDSS i & 7481 \AA & 1.3 & $0.649 \pm 0.005$ & 0.7 & (c) & 1.6\\
		SDSS z & 8931 \AA & 1.3 & $0.474 \pm 0.003$ & 0.8 & (c) & 5.2\\
		2MASS J & 1.24 \micron & 2.0 & $0.284 \pm 0.003$ & 1.7 & (d) & 13\\
		2MASS H & 1.66 \micron & 2.0 & $0.181 \pm 0.002$ & 1.9 & (d) & 32\\
		2MASS Ks & 2.16 \micron & 2.0 & $0.118 \pm 0.001$ & 1.9 & (d) & 22\\
		IRAC 3.6 & 3.6 \micron & 1.66 & $0.0528 \pm 0.0011$ & 3 & (e) & 2.8\\
        IRAC 4.5 & 4.5 \micron & 1.72 & $0.0361 \pm 0.0010$ & 3 & (e) & 8.9\\
		IRAC 8.0 & 8.0 \micron & 1.98 & 0.00 & 3 & (e) & 9.8\\
		MIPS 24 & 24 \micron & 6 & 0.00 & 5 & (f) & 7.6\\
        PACS 70 & 70 \micron & 5.8 & 0.00 & 7 & (g) & 22\\
		PACS 100 & 100 \micron & 6.9 & 0.00 & 7 & (g) & 16\\
		\hline
	\end{tabular}
    \begin{tablenotes}
    References calibration uncertainties: (a) \cite{2007ApJS..173..682M}. (b) Calculated from Table 1 of the SWIFT UVOT CALDB Release Note 16-R01: \url{https://heasarc.gsfc.nasa.gov/docs/heasarc/caldb/swift/docs/uvot/uvot_caldb_AB_10wa.pdf}. (c) SDSS DR12 Data Release Supplement: \url{http://www.sdss.org/dr12/scope/}. (d) \cite{2003AJ....125.2645C}. (e) IRAC Instrument Handbook: \url{https://irsa.ipac.caltech.edu/data/SPITZER/docs/irac/iracinstrumenthandbook/17/#_Toc410728305}. (f) MIPS Instrument Handbook: \url{https://irsa.ipac.caltech.edu/data/SPITZER/docs/mips/mipsinstrumenthandbook/42/#_Toc288032317}. (g) PACS Instrument \& Calibration Web Pages: \url{http://herschel.esac.esa.int/twiki/bin/view/Public/PacsCalibrationWeb#Photometer_calibration_in_scan_m}.  
    \end{tablenotes}
    \end{threeparttable}
\end{table*}

\section{Data reduction pipeline for SWIFT UVOT images of extended sources}
\label{sec: Data reduction}

Since the SWIFT UVOT instrument was designed to detect point sources, the standard data reduction pipeline is not suitable for extended sources. Therefore, we developed a new reduction pipeline that can easily be exploited to reduce the images for a large sample of galaxies. For the first time, we present here and in quite some depth our new pipeline, which will be made publicly available in the future.

Images were retrieved from NASA's HEASARC Archive, using the SWIFT Data Query form\footnote{\url{https://heasarc.gsfc.nasa.gov/cgi-bin/W3Browse/swift.pl}}. UVOT data are taken in 3 modes: Image mode, Event mode and Image\&Event mode. We only obtained the IMAGE data in the three UV filters: UVW2, UVM2 and UVW1. Our reduction pipeline uses several tasks from the specialized HEASoft software (version 6.17), which can be downloaded from the HEASARC website\footnote{\url{https://heasarc.nasa.gov/lheasoft/download.html}}. We created \textsc{python} scripts that perform these tasks in an automatic way to all images so that the pipeline can be applied to other galaxies very easily and efficiently. We furthermore implemented a pixel-by-pixel correction for coincidence loss (i.e. the systematic underestimation of flux when multiple photons arrive simultaneously at the photon counter, see section \ref{subsec: coiloss}). The rest of this section describes the different steps of our pipeline: aspect correction, creation of auxiliary maps, combination of separate frames, several corrections to the flux, combination of different epochs and calibration and aperture correction.

\subsection{Aspect correction}
\label{sec: aspect}
We start from the SKY maps retrieved from the archive. The astrometry of these individual maps is not exact and needs to be corrected for slight offsets. This so-called ``aspect correction'' happens in two steps, with the HEASoft task \texttt{uvotskycorr}\footnote{All tasks in \texttt{typewriter} font are HEASoft tasks, which will not explicitly be repeated in the remainder of the paper.}. In the first step, the aspect corrections are calculated and saved to a file. The task identifies stars in every frame and compares the detected sources with the USNO-B1.0 all-sky catalogue \citep{2003AJ....125..984M}. A correction is only calculated if at least three detected sources can be matched to sources in the catalogue. An aspect correction could not be found for all frames. This can be the case e.g. when there are not many (bright) stars in the field-of-view with a high enough signal-to-noise ratio (SNR). However, in our case, an aspect correction was found for most of the images. The other images are left out from further analysis. Subsequently, the aspect corrections are applied to the individual frames. The IDs of the images that were eventually used in the analysis are listed in Table \ref{tab: SWIFT_ID} in appendix \ref{app: ID}. Note that each of these ``images'' has several individual frames that need to be reduced separately.

\subsection{Creation of auxiliary maps}

At this point, three auxiliary maps must be created, which are needed in further steps of the reduction pipeline:

\begin{enumerate}[label=(\alph*),leftmargin=0.5\parindent]
\item Quality maps:

Quality maps contain flags associated with the goodness or badness of each detector pixel and are used to build exposure maps (see next step). We create quality maps for each image frame with the task \texttt{uvotbadpix}.

\item Exposure maps:

We generate exposure maps based on the aspect corrected frames and the created bad pixel masks with the task \texttt{uvotexpmap}. Exposure maps are used to normalize the frames in a later step (see section \ref{sec: combination}).

\item Large scale sensitivity maps:

Large scale sensitivity (lss) maps are created for every image frame with the task \texttt{uvotskylss} and are used in a later step to correct for the non-uniform sensitivity of the detector (see section \ref{sec: lss correction}).
\end{enumerate}

\subsection{Combination of separate frames and normalization}
\label{sec: combination}
In order to improve the SNR, the different sky frames should be combined. To co-add the frames it is important that they are perfectly aligned, i.e. that the aspect correction was calculated and applied correctly to each frame. Furthermore, at this stage, we can only add images that were observed more or less during the same period. Because the UVOT detectors suffer from sensitivity loss over the years, we need to correct for this in a later step (see section \ref{sec: zero point correction}). The later the images were taken, the larger this correction will be, so we need to separate images taken in different periods. In our case, we ended up with four summed images, one for each of the four years in which NGC\,628 was observed: 2007, 2008, 2013 and 2015. It is important to note that in this step we also sum the exposure maps and the large scale sensitivity maps. With the \texttt{ftappend} task, all frames are first appended into one single image, and with \texttt{uvotimsum} the different frames are then actually summed.

Subsequently, we normalize the total sky images so that the units are converted from counts to count rates (i.e. counts/s). This is required for the calculation of the correction factors in the following step (section \ref{sec: corrections}). Hereto, we divide the summed sky images by their corresponding summed exposure maps, using the task \texttt{farith}.

\subsection{Flux Corrections}
\label{sec: corrections}
\subsubsection{Coincidence loss correction}
\label{subsec: coiloss}

In the photon-counting detector employed in the UVOT, the incoming signal from each individual photon generates a splash of photons across multiple pixels on the CCD camera, with the centre of the splash giving positional information accurate to a sub-CCD-pixel scale for point sources. The detector has a readout speed (or frame time) of about 11\,ms in full-frame imaging mode \citep{Roming2005}, which may lead to a systematic underestimation of photon energy (so-called ``coincidence loss'' or ``pile-up'') when two or more photons arrive at a similar location on the detector within the same CCD readout interval. The detector is not sensitive above a certain photon count rate. Thus, for extended sources with higher count rates of incoming photons, it can happen that the simultaneous arrival of multiple photons is counted as one single photon, and/or that the detection of the incoming photons is misplaced by the centroiding algorithm.

\begin{figure}
	\includegraphics[width=\columnwidth]{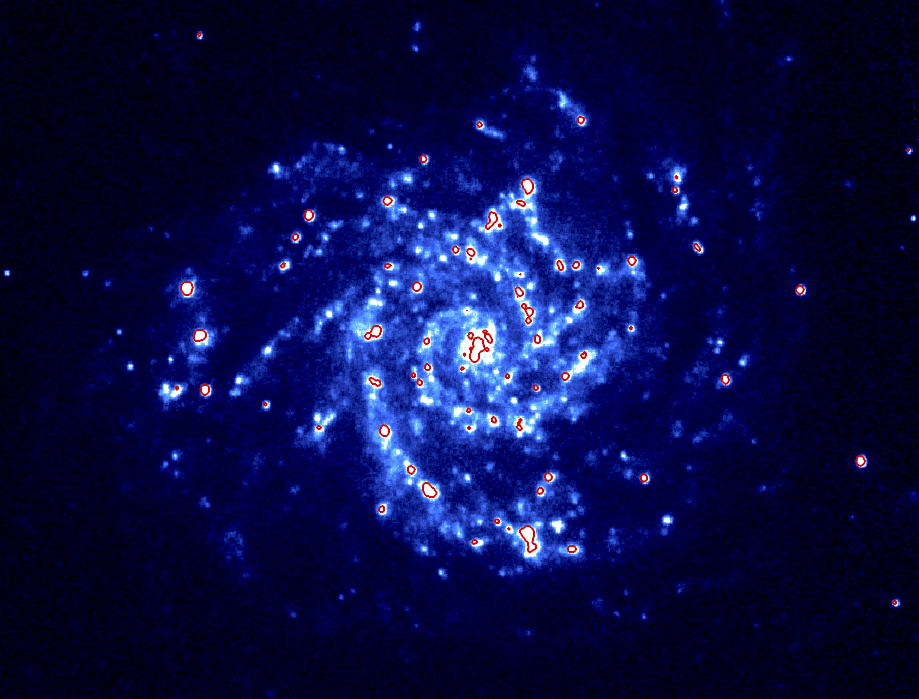}
    \caption{Co-added SWIFT UVW1 image from 2015 with contours indicating the regions with a count rate higher than 0.028\:counts/s/pixel, for which the coincidence loss becomes significant.}
    \label{fig: coiloss}
\end{figure}

We need to correct for this coincidence loss to recover the true incoming photon rate. Based on a detailed analysis of the UVOT response to different background models in \cite{2010MNRAS.406.1687B}, the coincidence loss correction has been shown to become non-negligible for regions with count rates higher than 0.007 counts/s/pixel for an unbinned pixel grid (with pixel size $0.5\arcsec\times0.5\arcsec$). We use $2\times2$ binned images (with pixel size $1\arcsec\times1\arcsec$), so the critical count rate is 0.028 counts/s/pixel. In Fig. \ref{fig: coiloss} we indicate the regions with a higher count rate in the co-added UVW1 image from 2015. Particularly in the spiral arms and the star-forming regions of NGC\,628 this threshold is easily exceeded, so a coincidence loss correction becomes inevitable.

Since the UVOT instrument was mainly designed to observe gamma-ray bursts and other high-energy sources, the existing calculations of coincidence loss corrections have been optimized for point sources. The correction factor is standardly derived from an aperture with a 5\arcsec\:radius around the point source. This choice of aperture size (with a radius of 5\arcsec) is driven by its minimal contribution from tails of the point spread function (PSF) \citep{2008MNRAS.383..627P}. The default procedure is not applicable to sources of angular extent larger than the size of the standard 5\arcsec\:aperture. \cite{2013MNRAS.431.2493K} present a method to correct for coincidence loss in extended sources in which they divide the galaxy image into regions of similar count rates, following isophotal contours. The coincidence loss correction factors are determined from representative 5\arcsec\:`test' apertures and applied to correct photon count rates in the corresponding regions. In \cite{2011AJ....141..205H} the resolved regions in M81, where coincidence loss corrections were required, were limited, and point-source corrections could be applied to a couple of individual regions. This approach is, however, not feasible for pixel-by-pixel studies. Therefore, we had to come up with a new strategy to calculate and apply coincidence loss corrections to extended sources. We adapted the correction technique for point sources of \cite{2008MNRAS.383..627P} to a pixel-by-pixel algorithm. Because of the photon splash that is created in the detector, one cannot simply calculate a correction factor for each individual pixel without taking into account the surrounding pixels. Therefore, we determine the coincidence loss correction factor within a $9\times9$ pixels sized box (or $81\arcsec^2$ in our case) centered on the pixel of interest, which covers an area equivalent to a $R=5\arcsec$ aperture region. The coincidence loss corrected count rate $C_{\textrm{coicorr}}$ (in counts/s) in a certain pixel is calculated as:
\begin{equation}
C_{\textrm{coicorr}} = C_{\textrm{obs}} \times f_{\textrm{coicorr}}
\end{equation}

\noindent where $C_{\textrm{obs}}$ is the observed count rate in that pixel and $f_{\textrm{coicorr}}$ is the coincidence loss correction factor determined within a $9\times9$ pixels sized box centered on that pixel. The detailed calculation of this correction and of the uncertainty on the corrected flux can be found in appendix \ref{app: coicorr}.

It must be mentioned at this point that our sliding-box correction method has its limitations. For very high flux rates (and thus large coincidence losses) the area around the star (or bright source) becomes dark because of the displacement of counts that was discussed earlier, which cannot be fixed with our method. Furthermore, when calculating the correction factor in a box that is larger than the source, the correction can be underestimated in the central regions of a bright source and overestimated around the bright region. However, in our case the typical coincidence loss corrections are relatively small compared to the overall uncertainties on the SWIFT fluxes (4.5-6\% on average). For the relevant regions (with pixel values exceeding the critical count rate of 0.028 counts/s/pixel) we find corrections between 0.5 and 3\%, with larger corrections only encountered in regions that were later masked as foreground stars. Moreover, since the images are rebinned in a later stage (see section \ref{sec: Data processing}) to pixels that are 7 times bigger, possible local over- and/or under-corrections are averaged out.

\subsubsection{Large scale sensitivity correction}
\label{sec: lss correction}

As explained before, the large scale sensitivity correction accounts for the non-uniform sensitivity of the detector (i.e. the sensitivity varies with large-scale changes of position on the detector). This step resembles the standard flat-fielding correction that is carried out for CCD cameras, but since a photon-counting instrument is insensitive to low-level CCD throughput variations, a traditional flat-field correction is not appropriate. For the details of the calculation of this correction, we refer the reader to section 5 of \cite{2010MNRAS.406.1687B}. The correction can be applied by dividing the coincidence loss corrected map by its corresponding lss map that was created earlier, using the \texttt{farith} task.

\subsubsection{Zero point correction}
\label{sec: zero point correction}

As mentioned before, the sensitivity of the detector decreases over the course of several years since the beginning of the SWIFT mission in January 2005. We need to correct for this sensitivity loss by applying a zero point correction ($f_{\textrm{zpcorr}}$), which can be calculated with the following quadratic function (from the SWIFT UVOT CALDB Release Note 15-03\footnote{\url{https://heasarc.gsfc.nasa.gov/docs/heasarc/caldb/swift/docs/uvot/uvotcaldb_throughput_03.pdf}\label{R15}}):
\begin{equation}
f_{\textrm{zpcorr}} = at^2 + bt + c
\end{equation}

\noindent with $t$ the time in years after January 1\textsuperscript{st} 2005 and the parameters $a$, $b$ and $c$ taken from Table 3 of the same Release Note\footref{R15}. Typical corrections are found to be between 1 and 20\% depending on the filter and the year in which the image was taken. The zero point corrected count rate $C_{\textrm{zpcorr}}$ (in counts/s) can then be calculated as:
\begin{equation}
C_{\textrm{zpcorr}} = \frac{C_{\textrm{coicorr}}}{f_{\textrm{zpcorr}}}
\end{equation}

The uncertainty on this correction is included in the total calibration uncertainty listed in Table \ref{tab:filters}.

\subsection{Combination of the different epochs}

Now that all corrections have been applied to the images, the four different summed images (from different years) can be combined. We therefore convert the units of the images from count rates back to the original counts, sum the images (and their corresponding exposure maps), and afterwards normalize the total image again by dividing by the total exposure map.

\subsection{Calibration and aperture correction}

In a final step, we convert the units of the combined image from count rates $C_{\text{tot}}$ (in counts/s) to flux densities (Jy): 
\begin{equation}
F (\text{Jy}) = C_{\text{tot}} \times f_{\text{cal}} \times 10^{23} 
\end{equation}

\noindent using the calibration factors $f_{\text{cal}}$ listed in the first column of Table 2 of the SWIFT UVOT CALDB Release Note 16-R01\footnote{\url{https://heasarc.gsfc.nasa.gov/docs/heasarc/caldb/swift/docs/uvot/uvot_caldb_AB_10wa.pdf}}.

It is very important at this point to realize that these calibration factors were determined based on apertures with a 5\arcsec\:radius, which only contain approximately 85\% of the total flux of a point source \citep{2010MNRAS.406.1687B}. Therefore, the calibration factors have been upscaled in such a way that using these factors in a 5\arcsec\:aperture around a point source results in the correct flux for that source. However, if we would use these factors directly to calibrate the flux in every individual pixel in the image, we would overestimate the flux. Therefore, we need to compensate for this by using the curve of growth (see Fig. 1 of \citealt{2010MNRAS.406.1687B}), which reaches a plateau for an aperture with a radius of about 30\arcsec. The total encircled energy in this ``maximum'' aperture, normalized to the encircled energy in a 5\arcsec\:aperture is exactly the aperture correction $f_{\text{apcorr}}$ that was applied to the calibration factors. To obtain the true pixel fluxes, we thus need to downscale them again:
\begin{equation}
F_{\text{corr}} (\text{Jy}) = F (\text{Jy}) / f_{\text{apcorr}}
\end{equation}

\begin{figure}
	\includegraphics[width=\columnwidth]{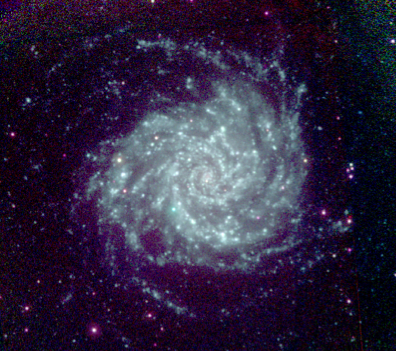}
    \caption{Colour image created by combining the three final reduced UV SWIFT images: UVW2 in blue, UVM2 in green and UVW1 in red.}
    \label{fig: SWIFT_color}
\end{figure}

\noindent (with $f_{\text{apcorr}}=$ 1.1279, 1.1777 and 1.1567 for the UVW2, UVM2 and UVW1 filters respectively). The final reduced images in the three filters were combined into a colour image, shown in Fig. \ref{fig: SWIFT_color}.

\section{Data processing with \textsc{pts}}
\label{sec: Data processing}
Further processing of the SWIFT images, as well as of all other waveband images (GALEX FUV/NUV, SDSS u/g/r/i/z, 2MASS J/H/Ks, IRAC\:3.6/4.5/8.0\,\micron, MIPS\:24\,\micron\ and PACS\:70/100\,\micron), is done by means of \textsc{pts}\footnote{\url{http://www.skirt.ugent.be/pts/}} (a \textsc{python} Toolkit for SKIRT), developed by Verstocken et al. (in prep). This toolkit is designed to post-process SKIRT radiative transfer simulations \citep{2003MNRAS.343.1081B,2011ApJS..196...22B,2015A&C.....9...20C}, but also to prepare images for SKIRT and for other purposes. The image processing includes subtraction of foreground stars/objects, correction for Galactic extinction, convolution, rebinning, background sky subtraction and uncertainty calculation. The code has been automated in such a way that it can be used for different images with very little manual intervention.

The removal of the foreground objects is performed in two steps. First, \textsc{pts} searches in the 2MASS All-Sky Catalogue of Point Sources II/246 \citep{2003yCat.2246....0C} for stars and in the HYPERLEDA I Catalogue VII/237 \citep{2003A&A...412...45P} for galaxies within the field-of-view of the image and tries to link these to actual sources in the image. Apart from stars and galaxies, the procedure will also detect other flux sources that do not belong to the target. In the second step, all detected sources are ``removed''. The pixel values in the detected regions are replaced by average local background values, determined by fitting a polynomial to the surrounding pixels and imposing a representative pixel-to-pixel noise variation, based on the standard deviation of the fitting. This approach ensures that the foreground objects are replaced by pixel values as close as possible to the (galaxy) emission in that particular region. When convolving the image in a later step, this results in a more realistic effect of those pixels on the surrounding areas, compared to when the pixels are simply masked. However, it must be noted that those ``artificial'' pixels were eventually excluded from the analysis (colour plots and spectral energy distribution (SED) fitting) in order to make sure we only use genuine measured fluxes.

Since we are interested in the attenuation properties of the dust in NGC\,628, we must correct for the extinction caused by dust in the MW. The MW extinction in the (CTIO) V-band for NGC\,628 is determined at $A_V = 0.188$ (obtained from the IRSA Galactic Dust Reddening and Extinction Archive\footnote{\url{https://irsa.ipac.caltech.edu/workspace/TMP_1vwZLo_26021/DUST/NGC628.v0001/extinction.html}}). Because this archive lacks extinction measurements in the GALEX and SWIFT bands, we convolve the MW extinction curve \citep{1989ApJ...345..245C} with the transmission curves of all filters, taken from the SVO Filter Profile Service\footnote{\url{http://svo2.cab.inta-csic.es/svo/theory/fps3/}}. The obtained Galactic foreground extinction $A_{\lambda}/A_V$ in all filters is given in Table \ref{tab:filters}. It should be noted that, in fact, this method will only give the correct MW extinction value if the background source has a flat response curve within every passband, which is of course not exactly true. We verified the effect of a non-flat spectrum by extinguishing the SED of the unattenuated stellar emission in NGC\,628 obtained with CIGALE (see section \ref{sec: CIGALE}) with the MW extinction curve. Comparing the attenuated and the unattenuated fluxes in every band results in the ``real'' MW extinction correction factors for all filters. The maximum differences between the correction factors found in this way and the average values that were used to correct the images for MW extinction (reported in Table \ref{tab:filters}) were added to the same table as an upper estimate of the uncertainty on the correction factors, and are negligible compared to the other uncertainties.

We convolve all images to the poorest resolution in our dataset, i.e. that of the PACS\:100\,\micron\ image, which has a PSF with a FWHM of 6.9\arcsec. We use the new 2018 Aniano kernels \citep{1538-3873-123-908-1218}\footnote{\url{https://www.astro.princeton.edu/~ganiano/Kernels/}}, which are rebinned to the same pixel grid as the image, normalized and aligned by \textsc{pts}. In view of performing a pixel-by-pixel analysis, we rebin all images to a common pixel grid with a pixel size of 7\arcsec $\times$ 7\arcsec, which is approximately equivalent to the resolution of the convolved images. In this way, the vast majority of the PSF is concentrated in a single pixel \citep{2012MNRAS.419.1833B,2015MNRAS.448..135B,2014A&A...567A..71V}. The physical size of each pixel is about 325\,pc.

Finally, the background emission needs to be subtracted from each image. We determine the sky background by randomly placing several apertures (typically about 100) with a radius of 4 times the FHWM of the PSF (about 28\arcsec\:in our case) in ``empty'' parts of the image where no sources were removed. The sky background is then estimated as the median of the mean flux densities in all these apertures, and is subtracted from the image.

For further analysis and for the SED fitting (section \ref{sec: CIGALE}), it is important to have an estimate of the uncertainties on the flux densities in every pixel of every image. There are several factors of uncertainty that need to be taken into account:
\begin{itemize}
\item Large scale variation in the background: This is computed as the standard deviation of the mean background values that were derived in the different apertures.
\item Pixel-by-pixel variation in the background: This is computed as the mean value of the standard deviations of the pixel values in the different background apertures. The total background noise (i.e. the square root of the quadratic sum of the large and small scale variations) in every image is given in Table \ref{tab:filters}.
 \item Calibration uncertainty: This is the uncertainty on the calibration of the instruments based on standard sources. Values are listed in Table \ref{tab:filters}.
\item Poisson noise: This accounts for the particle properties of light and the fact that the rate at which photons arrive at the detector is not exactly the same at every moment in time. The photons emitted at shorter wavelengths are less numerous, and therefore, the UV images will be more affected by this photon noise or Poisson noise. For this reason, we only calculate the Poisson noise for the GALEX and SWIFT images, starting from the original images in units of counts: Poisson noise = $\sqrt{N}$ with N the number of photons.\footnote{As described in \cite{2008MNRAS.383..383K} the uncertainty on the SWIFT measurements, in fact, does not follow a Poisson distribution but a binomial distribution due to the coincidence loss of the detector. However, for low count rates (< 0.05 counts/frame) the difference is negligible, as can be seen in their Fig. 1. Since the highest measured count rate in our images is 0.02 counts/frame, there will be no effect on the uncertainties.}
\end{itemize}

\noindent \textsc{pts} calculates these uncertainties and sums them quadratically (in every pixel) to compute a total uncertainty map on the flux densities.

As a final quality check of the SWIFT data, we compare the global galaxy flux in the SWIFT UVM2 and GALEX NUV bands, because these filters are centered around the same wavelengths. From the images we find a global flux of $(8.13 \pm 0.28) \times 10^{-2}$ Jy in the UVM2 band and of $(8.32\pm 0.24)\times10^{-2}$ Jy in the NUV band. These values are fully consistent, given the different filter properties.

\section{UV Colour-colour plots}
\label{sec: Colour plots}
\begin{figure*}
	\includegraphics[width=\textwidth]{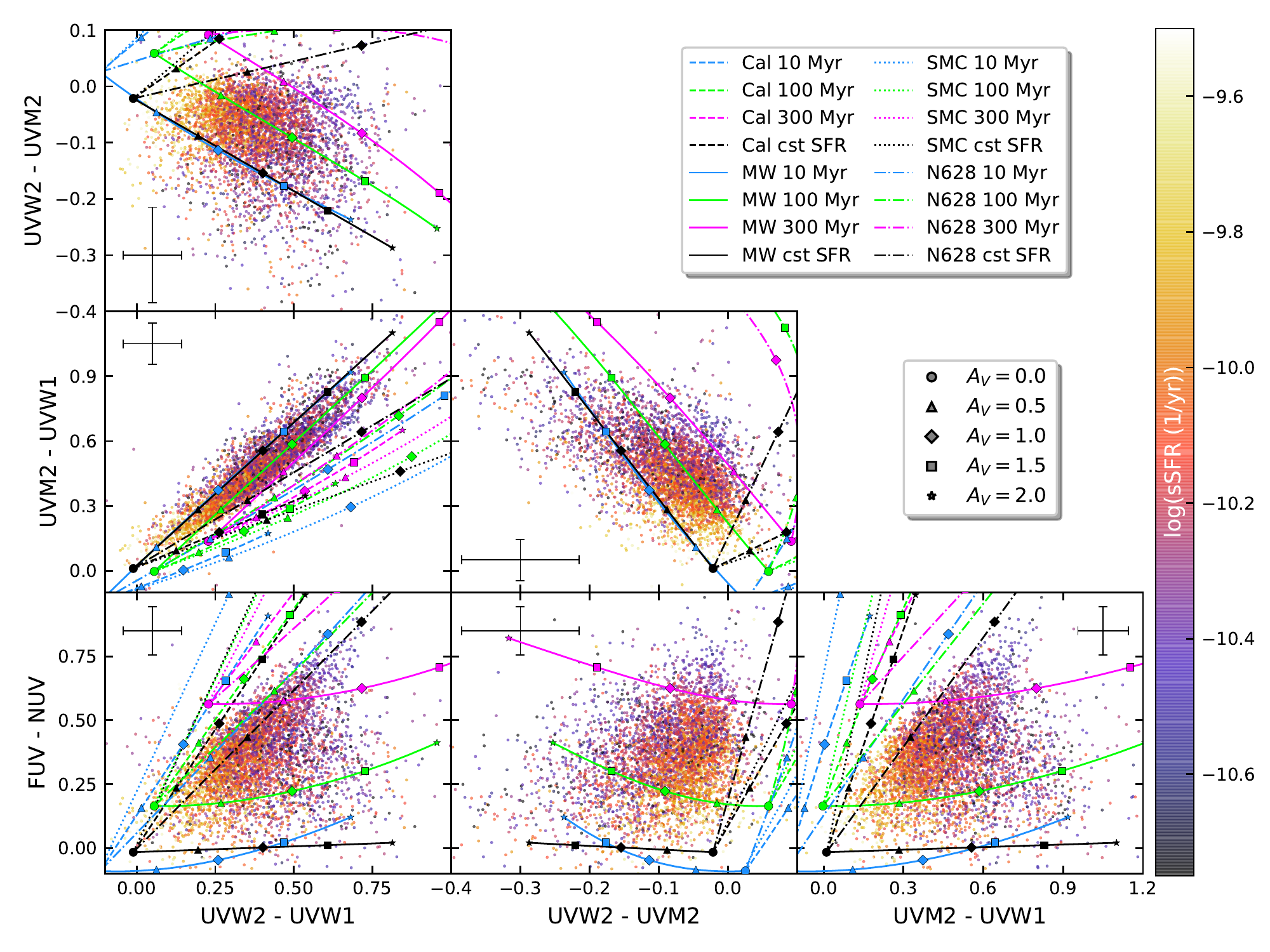}
    \caption{UV colour-colour plots. The dots represent the different pixels in NGC\,628 (5051 data points), whereas the lines are theoretical curves obtained by attenuating the spectra of SSPs (and a combined spectrum) with different levels of dust attenuation and different dust extinction and attenuation laws: the \citet{2000ApJ...533..682C} attenuation law (dashed), the MW extinction law from \citet{1989ApJ...345..245C} with $\textrm{R}_\textrm{V} = 3.1$ (solid), the SMC extinction curve from \citet{2003ApJ...594..279G} (dotted), and the median attenuation curve for NGC\,628 that we obtained from the \textsc{cigale} fitting (dash-dotted). A median error bar on the data points is shown. Some V-band attenuation levels are indicated ($A_{\textrm{V}}=0.0$ with circles, 0.5 with triangles, 1.0 with diamonds 1.5 with squares and 2.0 with stars). The data points are colour-coded according to their sSFR. (The stellar mass is derived from the IRAC\:3.6\,\micron\ image after correction for dust emission, while the SFR is derived from the FUV image, corrected for dust extinction.)}
    \label{fig: color}
\end{figure*}

A first analysis of the UV-data can be done through colour-colour plots, as given in Fig. \ref{fig: color}. They are a first step to compare models with observations and inform us about the stellar population and the level of attenuation in the galaxy. The dots represent the different pixels in the galaxy image with a SNR>5 in the five UV bands (resulting in 5051 data points), whereas the lines are theoretical curves obtained by attenuating the spectra of single stellar populations (SSPs) assuming different levels of dust attenuation and different dust extinction and attenuation laws (all shown in Fig. \ref{fig: att_curves}): the \cite{2000ApJ...533..682C} attenuation law (dashed lines), the MW extinction law from \cite{1989ApJ...345..245C} with $\textrm{R}_\textrm{V} = 3.1$ (solid lines), the SMC extinction curve from \cite{2003ApJ...594..279G} (dotted lines), and the median attenuation curve for NGC\,628 that we obtained from the \textsc{cigale} fitting (see section \ref{sec: results}; dash-dotted lines). The SSPs are taken from the library of \cite{2003MNRAS.344.1000B} with a \cite{2003PASP..115..763C} initial mass function, assuming a solar metallicity ($Z=0.02$) (and are also used in the modelling in section \ref{sec: CIGALE}). The different lines correspond to SSPs with different ages (10 Myr in blue, 100 Myr in green and 300 Myr in magenta). As we are looking at UV colours, we expect the light to be dominated by young stars of a few 100 Myr or younger, which is indeed confirmed by the data points in the plots. As will be explained at the end of this section, galaxies are generally better characterized by continuous-age rather than single-age populations. Therefore, we also added models (in black) for a stellar spectrum obtained by combining SSPs of different ages, assuming a constant SFR. In addition, some V-band attenuation levels are indicated ($A_{\textrm{V}}=0.0$ with circles, 0.5 with triangles, 1.0 with diamonds, 1.5 with squares and 2.0 with stars). Most regions in NGC\,628 seem to be consistent with an $A_{\textrm{V}}$ between 0 and 1.5. Finally, we colour-coded the data points according to the empirically derived sSFR in the corresponding pixel. It is clear that regions in the galaxy with higher sSFR exhibit younger stars, as expected.

The scatter plots compare four different UV colours: UVW2-UVM2, UVM2-UVW1, FUV-NUV and UVW2-UVW1. The UVW2-UVW1 and FUV-NUV colours are most sensitive to the UV slope of the attenuation curve, with the largest values for the FUV-NUV colour (at fixed stellar age and V-band attenuation $A_{\textrm{V}}$) observed for the steep SMC curve. Furthermore, there is a dependence on the bump strength (in the NUV band), which is reflected in smaller FUV-NUV colours for MW curves (which have a bump feature) compared to (bumpless) Calzetti curves, even though the UV slope (without bump) is steeper for the MW compared to the Calzetti law. Finally, the FUV-NUV colour also depends on the stellar age, especially for a MW-type dust extinction curve with a shallow UV slope, with larger values for older stellar populations. The UVW2-UVW1 colour probes the shape of the UV slope on opposite sides of the 2175\,\AA\ bump feature. Again, the largest values are obtained for the steepest curve (SMC) (at fixed stellar age and $A_{\textrm{V}}$). Most of the data points (for spatially resolved regions in NGC\,628) show a linear correlation in the FUV-NUV vs. UVW2-UVW1 diagram (see bottom left panel), following the trend of the median NGC\,628 curve. The theoretical models extincted with a MW curve are much flatter in this diagram (with almost constant FUV-NUV for a fixed stellar age), inconsistent with the observed trend for the resolved regions in NGC\,628, suggesting that the dust law in this galaxy is generally steeper than the MW curve (and/or has a less prominent bump feature).

The colours UVW2-UVM2 and UVM2-UVW1 are more sensitive to the bump feature with the UVM2 band coinciding with the position of the 2175\,\AA\ bump feature (see Fig. \ref{fig: uvot_filters}). Due to the longer tails of the UVW2 and UVW1 transmission curves, these colours do not perfectly measure the bump strength relative to the continuum at those wavelengths. The majority of resolved regions in NGC\,628 have a negative UVW2-UVM2 colour, implying that the observed UVM2 flux is lower than the UVW2 flux. Similarly, the positive UVM2-UVW1 colours imply that $\textrm{flux}_{\textrm{UVM2}} < \textrm{flux}_{\textrm{UVW1}}$. This dip in the flux around 2200\,\AA\ indeed suggests the presence of a bump feature in the attenuation curve around this wavelength. The scatter in the plots with UVW2-UVM2 colours (i.e., the top left and middle column panels) is dominated by the uncertainties on the SWIFT fluxes (over the limited dynamic wavelength range covered in UVW2-UVM2 colours), and does not help us much to distinguish between various theoretical models. Nevertheless, the data points in these plots seem to be covered mostly by a MW-type extinction curve, again a hint for a prominent bump feature. Likewise, the observed UVM2-UVW1 colours (in the bottom right and the middle row panels) can only be recovered by theoretical models assuming a MW-like bump strength (or a bump strength similar to the median NGC\,628 curve), also suggesting a relatively strong bump (within the error bars) for most regions in NGC\,628. The fact that most of the data points in the bottom right plot show a linear correlation following the trends of the steeper curves suggests that a flat UV slope, as in the MW curve, is not adequate to reproduce the data, and that a steeper slope is required (at least in a large part of the galaxy). Finally, the strongest correlation is observed in the UVM2-UVW1 vs. UVW2-UVW1 colour diagram. Due to the limited wavelength range probed in the colours on both axes of this plot, it is mostly sensitive to the bump strength (and not so much to the UV slope). The data points clearly coincide with the MW curves, meaning that a prominent bump feature must be present in the attenuation curve of NGC\,628.

An important note to make here is that stars have been forming throughout a galaxy's lifetime, and  usually do not form during a single burst of star formation. Therefore, in addition to simple single-age SSPs, we have also overlaid theoretical curves assuming a constant SFR (in black). In the most constraining colour diagrams (middle left, bottom left and bottom right panels) these curves agree relatively well with the observed colours (especially for the median NGC\,628 curve), indicating that this continuous age model is more adequate than the single-age SSPs to represent the resolved stellar populations in this galaxy. In summary, the colour plots suggest that the resolved regions in NGC\,628 are consistent with a UV slope steepness in between the MW and SMC dust extinction curves, and are best fitted with a MW-type bump strength. However, we need to be careful with our conclusions as the theoretical curves in these plots only represent a limited number of stellar ages, SFHs and dust attenuation curve shapes. Moreover, we are subject to degeneracies between the age of the stellar populations, the attenuation level and the shape of the attenuation curve, which cannot be ruled out from these simple colour-colour plots. Therefore, in the second part of this work, we perform a more accurate modelling of the underlying stellar population, and allow for more flexibility in the parameters describing the dust attenuation curve, using the SED fitting code \textsc{cigale}.

\section{Modelling dust attenuation with \textsc{cigale}}
\label{sec: CIGALE}
\subsection{Description and method}
\label{sec: cig_method}
In order to constrain the dust attenuation curve on resolved scales in NGC\,628, we use \textsc{cigale}\footnote{\url{https://cigale.lam.fr/}} (a \textsc{python} Code Investigating GALaxy Emission, \citealt{2009A&A...507.1793N, 2014ASPC..485..347R,2018arXiv181103094B}) to fit an SED to our multi-wavelength dataset. \textsc{cigale} can model the FUV to radio spectrum of galaxies and estimate their physical properties such as the SFR and SFH, the attenuation by dust, the dust luminosity and mass, the stellar mass, and many other physical quantities. The code relies on the dust energy balance principle: the energy that is emitted by dust in the IR matches the energy that was absorbed by dust in the UV-optical range. It uses a combination of several SSPs, SFHs and a flexible attenuation curve. A large grid of models is fitted to the data and the physical properties are determined through the analysis of the likelihood distribution. This is based on Bayesian statistics, in which the physical properties are estimated by weighting the models depending on their goodness-of-fit. This method also directly returns the uncertainties on the estimated quantities. \textsc{cigale} has been used extensively in a broad range of studies, such as the IRX-$\beta$ relation \citep{2012A&A...539A.145B}, attenuation properties of galaxies \citep{2011A&A...533A..93B,2012A&A...545A.141B,2013A&A...554A..14B,2017MNRAS.472.1372L}, SFR estimators \citep{2014A&A...561A..39B,2014A&A...571A..72B,2016A&A...591A...6B} and so on.

We use \textsc{cigale} to fit an SED for every individual pixel in NGC\,628 to the following bands: GALEX FUV/NUV, SWIFT UVW2/UVM2/UVW1, SDSS u/g/r/i/z, 2MASS J/H/Ks and IRAC\:3.6\,\micron/4.5\,\micron. In addition, we use the total infrared (TIR) luminosity in every pixel as an extra data point in the fitting (see later). We only fit those pixels with a sufficiently high SNR (>3) in all bands (3530 pixels in total). Since our main goal is to constrain the attenuation curve (and not the full SED), we want to limit the N-dimensional parameter space in the fitting as much as possible. Therefore, we restrict ourselves to the UV-optical-NIR wavelength range and solely estimate the dust attenuation properties and SFH. Dust emission can be represented by a simple fixed model as explained below. Going beyond the IRAC\:4.5\,\micron\ band would complicate the situation, and would require to fit dust emission templates, significantly increasing the number of SED models to be considered in the fitting. Finally, since the FIR images beyond 100\,\micron\ have a poorer resolution, we would not be able to do the fitting on resolved scales of 7\arcsec.

To avoid degeneracies between the age of stellar populations and the level of dust attenuation, which could potentially bias the inference of the dust attenuation curve parameters, we compensate the ``lack'' of FIR fluxes by adding the total dust luminosity as an extra constraint to the models, calculated in every pixel as prescribed by \cite{2013MNRAS.431.1956G} (see their Table 3):
\begin{equation}
S_{\textrm{TIR}} = 2.162 \times S_{24} + 0.185 \times S_{70} + 1.319 \times S_{100}
\end{equation}

\noindent with $S_{\textrm{TIR}}$ the TIR luminosity density in units of $\textrm{W/kpc}^2$, and $S_{24}$, $S_{70}$ and $S_{100}$ the luminosity density in the MIPS\:24\,\micron, PACS\:70\,\micron\ and PACS\:100\,\micron\ bands respectively. \cite{2001ApJ...550..195P} already used the total FIR as a constraint to model star formation and dust extinction in luminous starburst galaxies. This approach of fitting the UV/optical SED and including constraints on the dust emission directly from the TIR luminosity, referred to as IR luminosity-constrained SED fitting (SED+LIR fitting), is also used by and described in \cite{2018ApJ...859...11S}. In this way the IR luminosity is treated as another SED ``flux'' point that directly constrains the SFR and the attenuation curve without fitting the shape of the IR SED, which significantly reduces the N-dimensional parameter space and consequently the computing time.

Important physical components and processes that are included in the models are the SFH, the stellar photospheric emission, the nebular emission (including continuum and recombination lines) and the dust attenuation and emission. For the SFH we use the \textsc{sfhdelayedflex} module \citep{2017A&A...608A..41C} which takes into account that the star formation did not start suddenly but more gradually, and which allows for a recent enhancement or decline of the SFR. For the stellar spectrum the module \textsc{bc03} is adopted, which is built on the SSP library of \cite{2003MNRAS.344.1000B}, with a \cite{2003PASP..115..763C} initial mass function and assuming a solar metallicity ($Z=0.02$). The same SSPs were used in the empirical colour analysis in section \ref{sec: Colour plots}. The nebular emission is modelled based on nebular templates from \cite{2011MNRAS.415.2920I}. For the dust attenuation we use the \textsc{dustatt\_calzleit} module which is based on the \cite{2000ApJ...533..682C} starburst attenuation curve. The global slope of the curve can be modified by multiplying it by a power law function $A(\lambda) \propto \lambda^{\delta}$, and the UV bump is modelled as a Lorentzian-like Drude profile described by 3 parameters: its central wavelength (which is kept fixed at 2175\,\AA), its width (or FWHM, fixed at 350\,\AA) and its amplitude $B$. In this parameterization, a \cite {2000ApJ...533..682C} curve corresponds to $\delta=0$ and $B=0$, while the MW extinction curve is approximately equivalent with $\delta \approx -0.15$ and $B=3$. It is important to note that all stellar populations are attenuated with the same law and same amount of attenuation, so there is no differential reddening between younger and older stellar populations. This is, of course, a simplification as a differential (age-dependent) extinction will lead to changes in the effective (observed) attenuation law due to the fact that young stars suffer a higher attenuation than older stars since they are still enshrouded in their birth clouds \citep{2007MNRAS.375..640P,2000ApJ...539..718C}. The effect of this assumption on the results will be verified in section \ref{sec: test}. Also the nebular emission lines are attenuated with the same curve. Finally, the dust emission is modelled by the simple \textsc{dale2014} module using the dust templates of \cite{2014ApJ...784...83D} (with fixed slope $\alpha = 2.0$). The parameter values for the SSP library, the nebular emission and the dust emission are kept fixed, while we let the code fit the parameters of the SFH and the dust attenuation. All parameter ranges used in the fitting are listed in Table \ref{tab: cig_params} in appendix \ref{app: cigale_par}. \textsc{cigale} makes use of flat priors. Combining all possible parameter values the code computed and fitted 6,561,000 models to each pixel.

\subsection{Fitting results}
\label{sec: results}
\begin{figure*}
	\includegraphics[width=14cm]{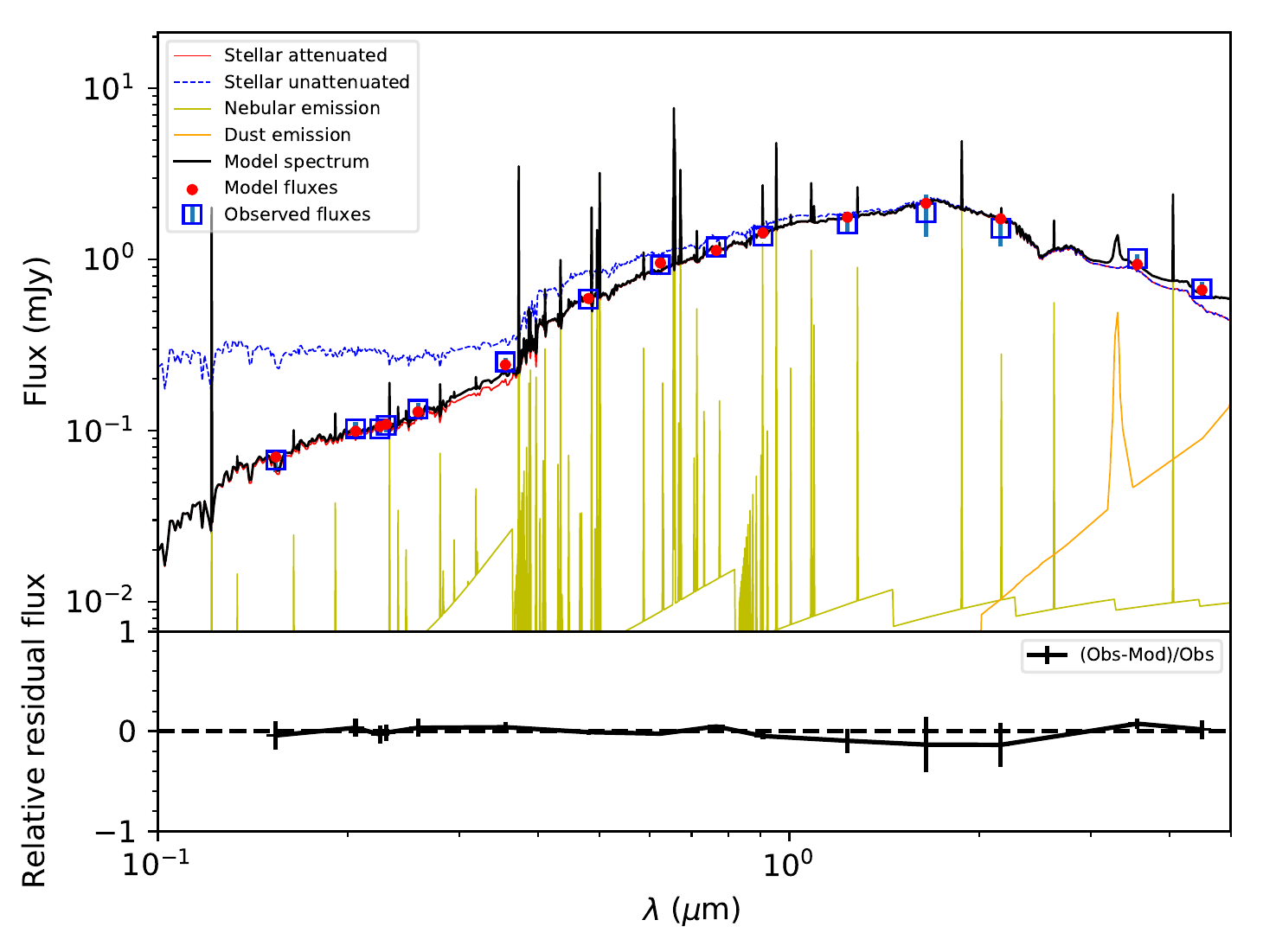}
    \caption{Top: Best-fitting model SED for pixel 261,258 obtained with \textsc{cigale}. Bottom: Relative residual flux between model and observations. The observed fluxes are fitted very well (within the error bars).}
    \label{fig: SED_highSFR}
\end{figure*}
\begin{figure*}
	\includegraphics[width=\textwidth]{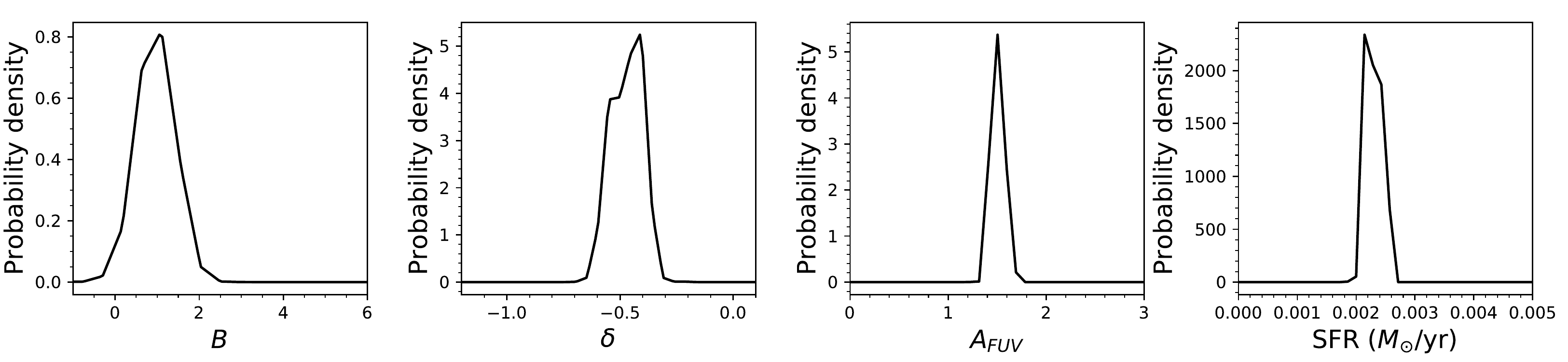}
    \caption{Probability distribution functions of several physical properties obtained with \textsc{cigale} for the fit corresponding to pixel 261,258: the UV bump amplitude $B$, the slope $\delta$, the FUV attenuation $A_{\text{FUV}}$ and the SFR are all well constrained.}
    \label{fig: pdfs}
\end{figure*}
\begin{table*}
	\centering
	\caption{Bump and slope values obtained with \textsc{cigale} for pixel 261,258 and median values for NGC\,628 for the ``standard'' case explained in section \ref{sec: cig_method}, and for the different tests described in section \ref{sec: test}: leaving out the SWIFT fluxes, changing the SFH, adding differential reddening, and fitting the global SED.}
	\label{tab: results}
    \begin{threeparttable}
	\begin{tabular}{lllllll}
		\hline
		parameter & pix 261,258 & NGC\,628* & No SWIFT*& \textsc{sfh2exp}* & diff. red.* & NGC\,628 global\\
		\hline
		$B$ & 0.85 $\pm$ 0.48 & 2.31 $\pm$ 0.95 & 2.05 $\pm$ 1.15 & 2.93 $\pm$ 1.08 & 1.87 $\pm$ 0.91 & 2.24 $\pm$ 1.16\\
    	$\delta$ & -0.47 $\pm$ 0.07 & -0.37 $\pm$ 0.10 &-0.32 $\pm$ 0.10 & -0.35 $\pm$ 0.12 & -0.19 $\pm$ 0.10 & -0.55 $\pm$ 0.15\\
		\hline
	\end{tabular}
    \begin{tablenotes}
    *$A_\textrm{V}>0.2$
    \end{tablenotes}
    \end{threeparttable}
\end{table*}

\begin{figure*}
	\includegraphics[width=14cm]{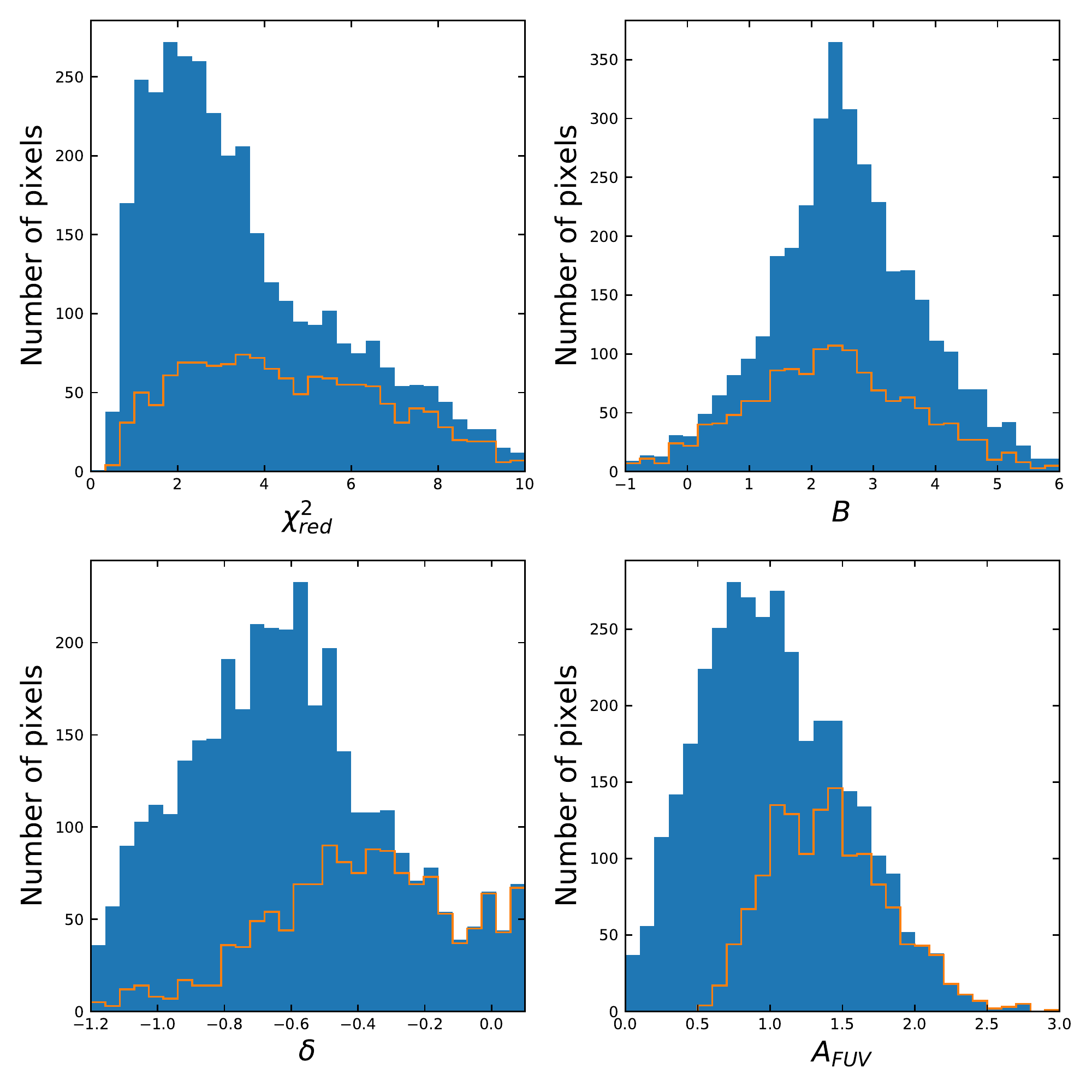}
    \caption{Histograms of the reduced $\chi^2$ of the fitting, the bump strength $B$, the slope $\delta$  and the FUV attenuation $A_{\text{FUV}}$, obtained with \textsc{cigale}. We overplot the histogram (in orange) for those pixels with $A_{\text{V}}>0.2$.}
    \label{fig: hist}
\end{figure*}

In Fig. \ref{fig: SED_highSFR} an example of a fitted SED can be found (in this case for pixel 261,258 in the central spiral arm, indicated with a red cross on Fig. \ref{fig: maps}). We can see that overall the observed data points are fitted very well. This is also true for the other pixels, which is reflected in the residual maps and histograms in Fig. \ref{fig: UV_maps} in appendix \ref{app: figures}. Residuals mostly lie within the typical uncertainties on the observed flux. We refer the reader to appendix \ref{app: figures} for a more detailed discussion on the fitting results in the different wavebands. In Fig. \ref{fig: pdfs} we show the probability distribution functions of some physical properties for pixel 261,258. The slope $\delta$, the FUV attenuation $A_{\text{FUV}}$ and the SFR are constrained reasonably well, while the UV bump amplitude $B$ has a somewhat wider distribution. This is also confirmed by the mock tests described in appendix \ref{app: mock}. Table \ref{tab: results} gives the bayesian bump and slope values obtained for this pixel.

We learn from the histograms in Fig. \ref{fig: hist} that there is a large variation in bump strengths and slopes among the different resolved regions in NGC\,628. As explained in the previous section, the slope represents the deviation from a Calzetti curve. Negative $\delta$ values mean that the curve is steeper than the Calzetti curve. We find a median bump strength of 2.31 with a spread of 1.33 and typical uncertainties around 0.95, and a median slope of -0.37 with a spread of 0.28 and typical uncertainties around 0.10, for regions with a V-band attenuation $A_{\text{V}}>0.2$ (see Table \ref{tab: results}). The uncertainties on these parameters are substantial, but from our mock data analyses (described in appendix \ref{app: mock}) it is clear that both parameters are constrained reasonably well, especially for regions with $A_{\text{V}}>0.2$. For pixels with a very low dust attenuation it is more difficult to get meaningful constraints on the attenuation curve. Therefore, in the remainder of this paper we will focus on regions with $A_{\text{V}}>0.2$ because the fitting results are more reliable for these regions.

\begin{figure}
	\includegraphics[width=\columnwidth]{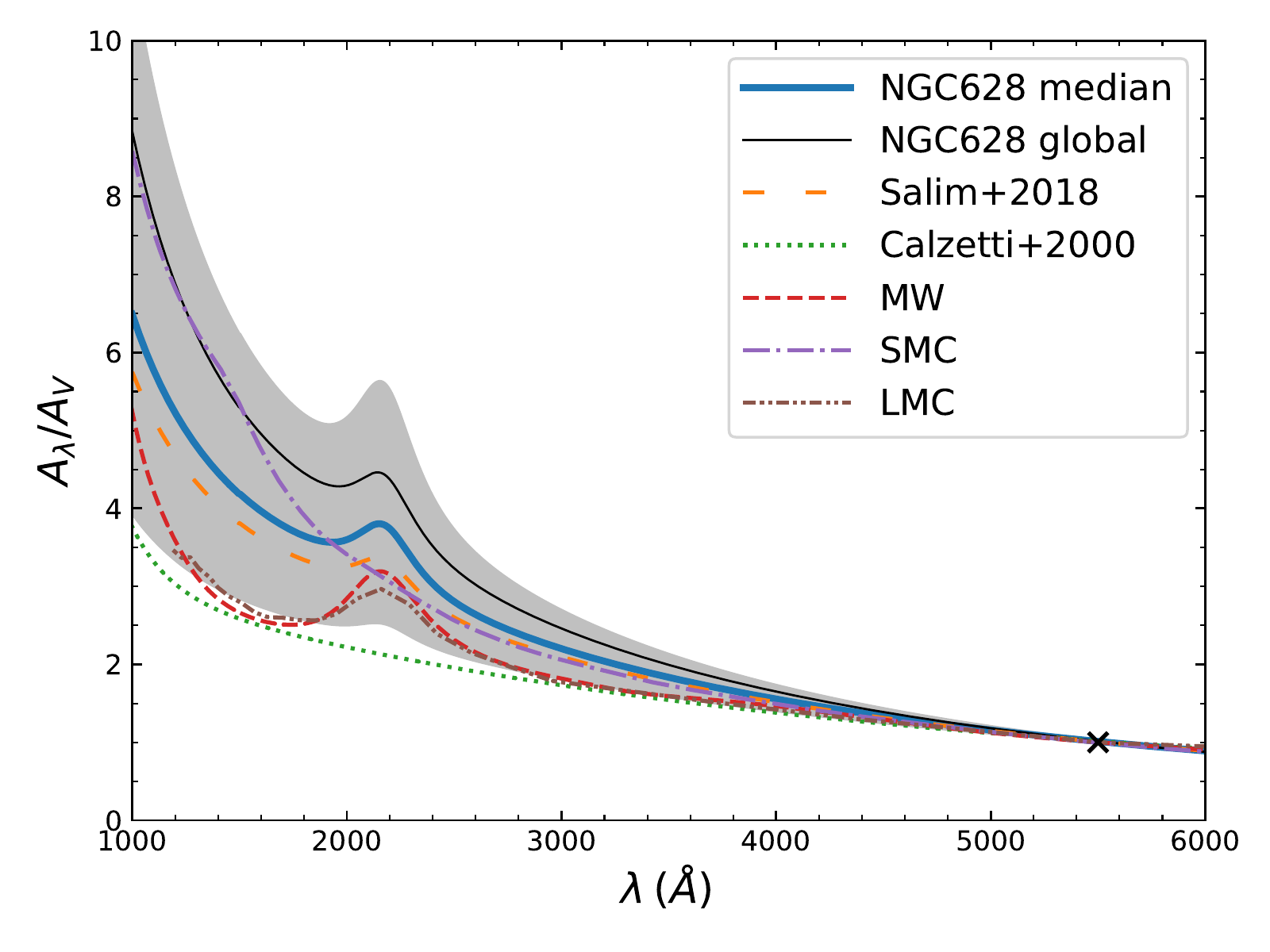}
    \caption{Extinction/attenuation curves for different galaxies. The thick blue line is the median attenuation curve that we obtained for NGC\,628 (for $A_{\text{V}}>0.2$ regions). The region between the 16\textsuperscript{th} and 84\textsuperscript{th} percentile is shaded. The black line is the global attenuation curve of NGC\,628. The orange dashed line corresponds to the average attenuation curve from \citet{2018ApJ...859...11S} for star-forming galaxies with $9.5 < \textrm{log} (M_*/M_{\odot}) \leq 10.5$ (with $B=1.73$ and $\delta=-0.3$). The green dotted line is the standard \citet{2000ApJ...533..682C} curve. The red dashed line is the average MW extinction curve from \citet{1989ApJ...345..245C}. Both the SMC (violet dash-dotted) and LMC (brown dash-dot-dotted) extinction curves are taken from \citet{2003ApJ...594..279G}.}
    \label{fig: att_curves}
\end{figure}

We plotted our ``median'' attenuation curve for NGC\,628 (i.e. an attenuation curve with the median bump and slope values stated above, with $A_{\text{V}}>0.2$) in Fig. \ref{fig: att_curves}, together with some other relevant curves: the MW extinction curve (from \citealt{1989ApJ...345..245C}), the SMC and LMC extinction curves (from \citealt{2003ApJ...594..279G}), the \cite{2000ApJ...533..682C} curve, and the average curve for star-forming galaxies with $9.5 < \textrm{log} (M_*/M_{\odot}) \leq 10.5$ found by \cite{2018ApJ...859...11S} (with $B=1.73$ and $\delta=-0.3$, see their Table 1). The shaded region in the plot represents the attenuation curves between the 16\textsuperscript{th} and 84\textsuperscript{th} percentile of bump and slope values (for $A_{\text{V}}>0.2$ regions). From this we conclude that the median attenuation curve of NGC\,628 has a somewhat shallower slope than the SMC curve ($\delta \approx -0.45$) and a bump that is somewhat smaller compared to the MW curve ($B=3$). This median curve was also used in the colour plot analysis in Fig. \ref{fig: color}.

\subsection{Testing of our model}
\label{sec: test}
In this section, we verify whether we are able to constrain the attenuation properties with \textsc{cigale} and whether the SWIFT data aid in this. We also test how our model assumptions on the SFH and the differential reddening affect the modelling output. Finally, we compare the fitting on resolved scales to a fit of the global galaxy SED.

In order to proof that our \textsc{cigale} model is able to constrain the physical galaxy properties we performed a mock data analysis. Input data generated from model SEDs for which we know the true parameters are fitted with the same bayesian approach as before. From this mock testing we conclude that the model is able to constrain all properties within reasonable uncertainties for regions with a V-band attenuation $A_{\text{V}}>0.2$. For more details on the method and the results of this mock analysis we refer the reader to appendix \ref{app: mock}.

To verify the usefulness of the SWIFT bands in constraining the dust attenuation curve, we repeat the exact same fitting procedure, but without the 3 SWIFT fluxes. The results for the median bump and slope, and their uncertainties are in line with the values found before (see Table \ref{tab: results}). However, we find that the median relative uncertainty on the bump strength reduces from 56\% to 41\% when adding the SWIFT data. Although it remains hard to constrain the bump feature, the SWIFT data clearly assist in this. Since the uncertainties on the SWIFT fluxes (4.5-6\% on average) are somewhat larger than the GALEX NUV uncertainty (around 4\%), the SWIFT bands are given a smaller weight in the \textsc{cigale} fitting compared to the GALEX NUV. We found that ``artificially'' reducing the SWIFT uncertainties so that the UVM2 uncertainty corresponds to that on the GALEX NUV flux, further reduces the median relative uncertainty on the bump strength down to 34\% (i.e. 22\% lower compared to the case without SWIFT data). We thus conclude that when the uncertainties on the SWIFT data are lower, they are more helpful in constraining the dust attenuation curve. We expect, for example, that for galaxies with longer exposures (which reduces the Poisson noise) the SWIFT data will be even more valuable in constraining the dust attenuation curve on resolved scales.

To check the influence of the SFH assumptions on the results, we did the same modelling as before but this time using the module \textsc{sfh2exp}. The SFH is parameterized with two decaying exponentials: the first exponential models the long-term star formation, which has formed the bulk of the stellar mass, while the second one models the most recent burst of star formation. The obtained results (see Table \ref{tab: results}) are not very different (within the error bars). However, our mock data analysis (see appendix \ref{app: mock}) showed that the SFH does have a (limited) influence on the results, especially for regions with lower $A_{\text{V}}$. This can indicate that the very steep slope values that we obtained in low $A_{\text{V}}$ (<0.2) regions are possibly biased by the assumptions on the SFH. However, we argue that the delayed and flexible SFH is more adequate to model the resolved regions in NGC\,628 compared to the double exponentially declining SFH. The delayed SFH with a final burst/quench is better adapted because it can more easily represent the more diverse SFHs we expect at a local scale in the galaxy disc: spiral arms had a recent episode of star formation, whereas star formation in interarm regions has fallen. This cannot be obtained with a double exponentially declining SFH. Therefore, we retain the \textsc{sfhdelayedflex} module.

We also tested whether introducing a differential reddening (i.e. using a different attenuation level for young and old stellar populations due to the extra attenuation for young stars from dust in their birth clouds) changes anything to the results presented in this work. To this aim, we have repeated the modelling of NGC\,628 assuming the same dust attenuation curve for young and old stellar populations, but accounting for a different level of dust attenuation, i.e. $E(B-V)_{\text{old}}=0.44\:E(B-V)_{\text{young}}$ or $A_{\text{V,young}}/A_{\text{V,old}}=2.27$, consistent with \citet{2000ApJ...533..682C}, for stars younger than 10\,Myr. The median results of this test can be found in Table \ref{tab: results}. The bump strength is not very different (within the error bars), but the attenuation curve is shallower. This is to be expected as explained in \citet{2018ApJ...859...11S}: the effective attenuation curve (i.e. when ``combining" young and old stars, assuming the same level of attenuation for both populations, as assumed throughout this paper) will be steeper (by $\Delta\delta=-0.2$) than the intrinsic ``individual" curves of the young and old populations (as obtained in this test when using a differential reddening), because the highly attenuated young stars dominate at shorter wavelengths. This is indeed what we observe. However, in the interest of making the simplest assumptions on the geometry and levels of attenuation for different stellar populations, we prefer to use the same level of attenuation for young and old stars.

Finally, we tested whether a fit of the global galaxy SED results in similar attenuation curve parameters as the median values obtained from the pixel-by-pixel fits. We used the same \textsc{cigale} setup as before, but fitting the global fluxes of the galaxy, measured in an aperture that includes all pixels that were fitted before and using the same masks for the foreground stars. The results of this run (see Table \ref{tab: results}) are close to the median values from the pixel-by-pixel fitting (within the error bars). Thus, for NGC\,628, an investigation of the attenuation curve on resolved scales gives similar results to our finding on a global galaxy scale. This is very promising for our ongoing project (DustKING) studying the global dust attenuation curves of a statistical sample of nearby galaxies (Decleir et al., in prep., and see section \ref{sec: comparison}). The global attenuation curve of NGC\,628 was added to Fig. \ref{fig: att_curves}. We also found that this global run significantly benefits from the SWIFT data points. Leaving out the SWIFT fluxes results in a median relative uncertainty of 124\% on the bump strength (which is even consistent with no bump) and 71\% on the slope (compared to 52\% and 27\% when adding SWIFT data, respectively). We thus conclude that the SWIFT data are a great added value to constrain the attenuation curve on global galaxy scales.

From the global SED fit of NGC\,628 we obtain a total $\text{SFR} = 1.30 \pm 0.16 \:M_{\odot}/\textrm{yr}$ and a stellar mass of $\textrm{log} (M_*/M_{\odot}) = 10.16 \pm 0.04$, which is in reasonable agreement with values found in the literature ($\text{SFR} = 1.07\:M_{\odot}/\textrm{yr}$; $\textrm{log} (M_*/M_{\odot}) = 9.821$ -- \citealt{2018arXiv180904088H}). \cite{2018ApJ...859...11S} studied the dust attenuation curves of 230,000 individual galaxies in the local Universe, in relation to the position of galaxies on the sSFR-$M_*$ diagram. In their Fig. 3, NGC\,628 is positioned very close to the line of median sSFR for star-forming galaxies in this diagram  (i.e. the ``main sequence'' of star formation). From this figure, we infer a slope between -0.4 and -0.3 and a bump between 1.0 and 1.5 for galaxies with a sSFR and stellar mass similar to NGC\,628. We obtain a global SED with a bump strength that is consistent within the error bars, but a steeper slope, possibly because we are using the same attenuation level for young and old stars, whereas \cite{2018ApJ...859...11S} assume a differential reddening (see the discussion higher-up).

\section{Discussion of the results}
\label{sec: Discussion}
\subsection{Resolved stellar and dust parameters}
\label{sec: resolved}
\begin{figure*}
	\includegraphics[width=\textwidth]{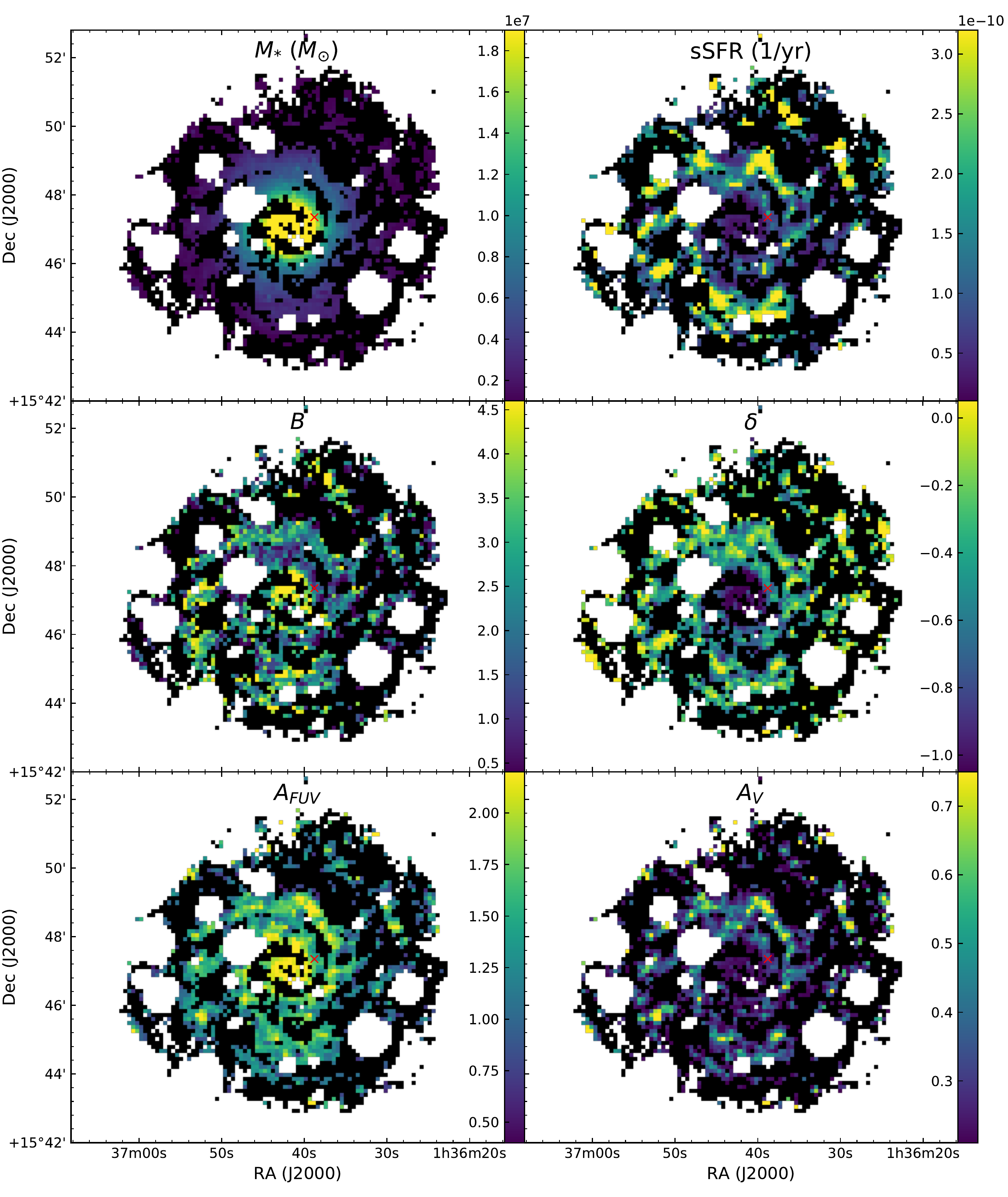}
    \caption{Maps of the stellar mass $M_{*}$ (in $M_{\odot}$), the sSFR (in 1/yr), the UV bump strength $B$, the attenuation slope $\delta$, the FUV attenuation $A_{\text{FUV}}$ and the V-band attenuation  $A_{\text{V}}$ for NGC\,628, all derived from the \textsc{cigale} fitting. Pixels with $A_\textrm{V}\leq 0.2$ are shown in black. The red cross indicates pixel 261,258 in the central spiral arm, for which the fitting results can be found in Table \ref{tab: results}.}
    \label{fig: maps}
\end{figure*}

In Fig. \ref{fig: maps} we show the distribution of the parameter values obtained with \textsc{cigale} across the disc of NGC\,628. The stellar mass ($M_{*}$) map shows a smooth distribution of the stars, with most stars located in the central regions of the galaxy, and a less pronounced increase in stellar mass surface density in features corresponding to the galaxy's spiral arms. The sSFR map reveals the location of the star forming regions in the spiral arms of the galaxy. The maps of the bump strength $B$ and the slope $\delta$ show the spatial variations of these parameters within the galaxy. Interestingly, we find that the centre of the galaxy is characterized by very steep slopes and strong bumps. However, the V-band attenuation ($A_{\text{V}}$) map shows little dust in the central region. We argue that the bulge of the galaxy is dominated by older stellar populations and that our assumed attenuation curve (which is based on the Calzetti starburst law) has the property to give low attenuation values in the NIR. In other words, the curve may underestimate dust heating from old stars. \textsc{cigale} will try to compensate the ``shortage'' of dust emission in the IR with very steep attenuation curves (and thus high $A_{\text{FUV}}$ values). Furthermore, the reduced $\chi^2$ values in the central region are relatively high compared to other galaxy regions, which confirms that the central bulge of the galaxy is harder to fit with our model set-up.
The interpretation of spatial variations in the bump and slope maps will be further discussed in section \ref{sec: Trends}.

The $A_{\text{V}}$ map shows that the V-band attenuation is below 1 across the galaxy disc, in agreement with the nearly transparent discs for galaxies viewed face-on as inferred from dust radiative transfer models \citep[e.g.][]{1999A&A...344..868X,2007A&A...471..765B,2014A&A...571A..69D,2014MNRAS.441..869D,2018A&A...616A.120M}.

\subsection{Trends between the shape of the dust attenuation curve and other galaxy properties}
\label{sec: Trends}

\begin{figure}
	\includegraphics[width=\columnwidth]{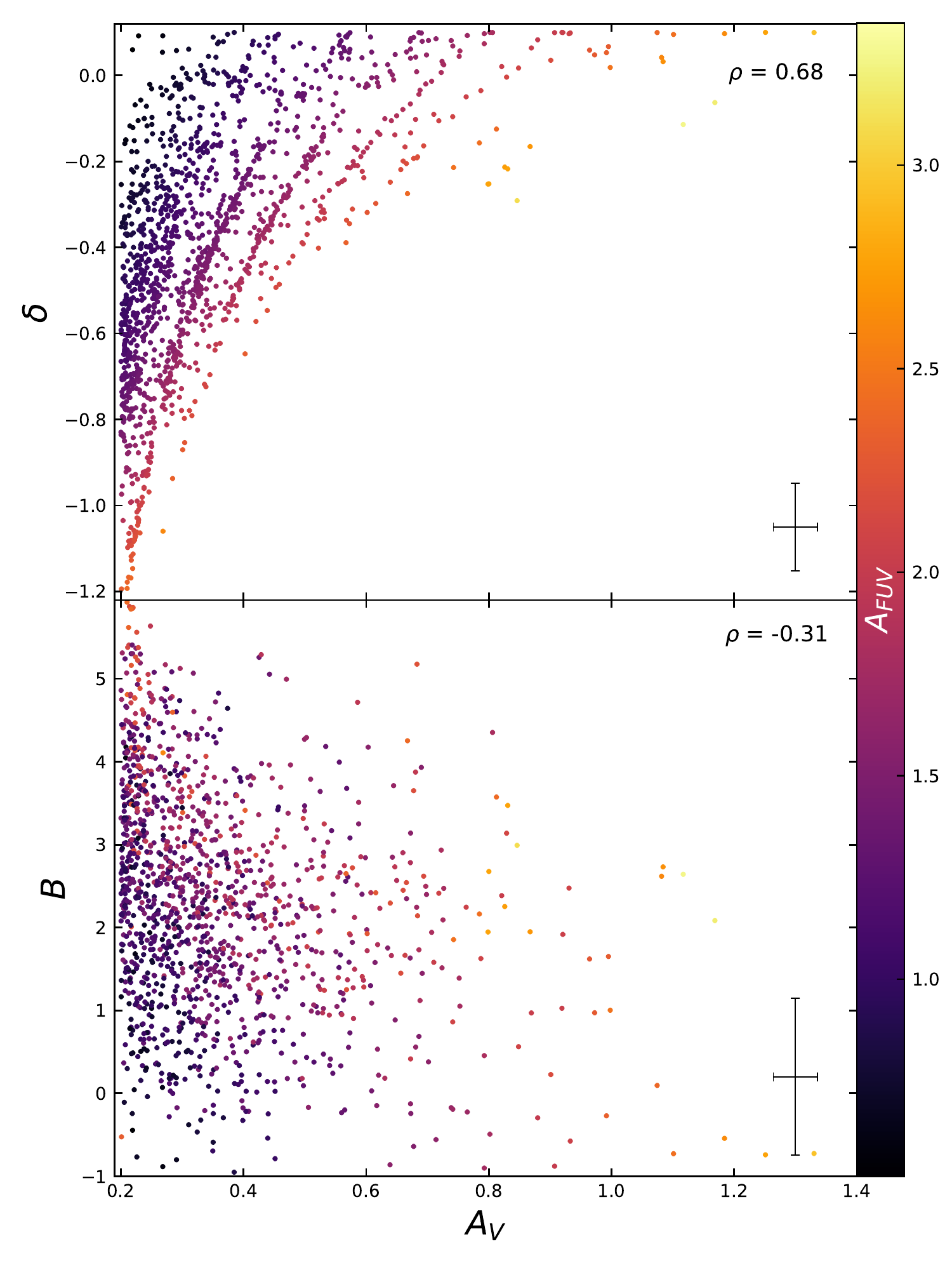}
    \caption{The attenuation curve slope $\delta$ and the bump strength $B$ vs. the V-band attenuation $A_{\text{V}}$, colour-coded with the FUV attenuation $A_{\text{FUV}}$. Typical error bars as well as the Spearman correlation coefficient $\rho$ are indicated.}
    \label{fig: AV}
\end{figure}

In this section, we attempt to explain the observed variety in the shape of the dust attenuation curve within the galaxy by looking at possible trends between the attenuation curve properties and other galaxy properties. In the top panel of Fig. \ref{fig: AV}, we see a correlation between the slope of the attenuation curve and the V-band attenuation $A_{\text{V}}$\footnote{Note that it is also possible to use the colour excess $E(B-V)$ as a measure for the amount of attenuation. However, in this case the trends are more difficult to interpret and to compare to other studies which use $A_{\text{V}}$, e.g. \citet{2017MNRAS.466.4540H}, \citet{2018ApJ...859...11S}, and \cite{2018ApJ...869...70N}.}, with shallower slopes corresponding to higher $A_{\text{V}}$ values. A similar trend was inferred by \citet{2017MNRAS.466.4540H}, \citet{2018ApJ...859...11S}, \citet{2018MNRAS.475.2363T} and \cite{2018ApJ...869...70N}, and interpreted in light of models presented in \citet{2013MNRAS.432.2061C}. More specifically, in regions with higher levels of dust attenuation, dust absorption dominates over scattering processes which results in a flattening of the dust attenuation curve. This suggests that the relative geometry of stars and dust dominates variations in the shape of the dust attenuation curve as opposed to alterations in intrinsic grain species. \cite{2018ApJ...869...70N} also conclude that the slope primarily depends on the star-dust geometry, based on simulations with 3D Monte Carlo dust radiative transfer calculations. They found that a flatter attenuation curve can be attributed to a mixed geometry with a larger fraction of unobscured young stars (decreasing the UV attenuation) and/or more obscured old stars (increasing the optical attenuation).

\begin{figure}
	\includegraphics[width=\columnwidth]{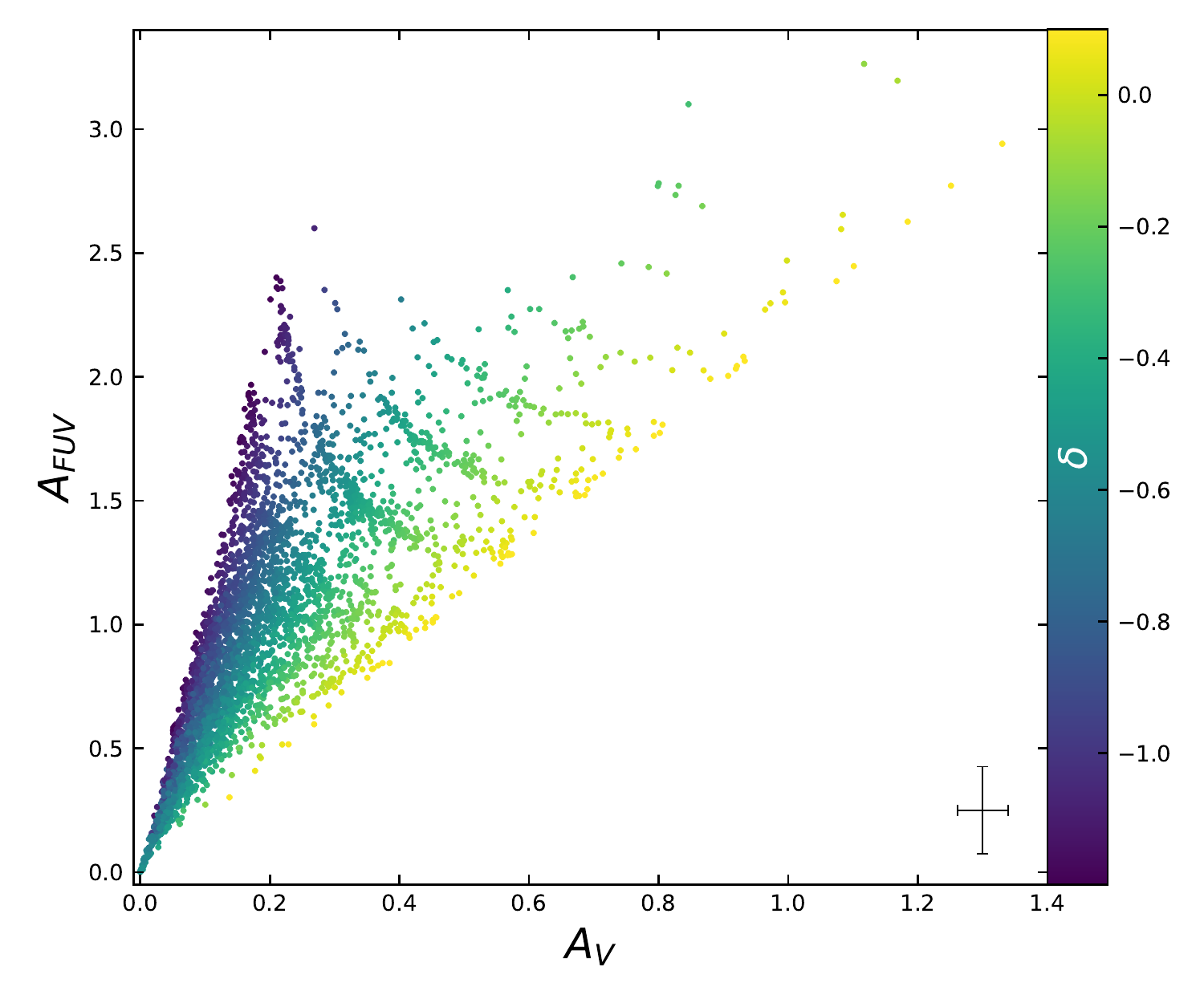}
    \caption{FUV attenuation $A_{\text{FUV}}$ vs. V-band attenuation $A_{\text{V}}$, colour-coded with the attenuation slope values. Typical error bars are indicated. The ``upper'' data points (in purple) correspond to the steepest slopes, whereas the ``lower'' points (in yellow) correspond to the shallowest slopes.}
    \label{fig: AFUV_AV}
\end{figure}

In the same plot the dependence of the slope on the FUV attenuation can be seen: at a fixed $A_{\text{V}}$ the slope tends to be steeper for higher $A_{\text{FUV}}$. This trend was also found by \citet{2018ApJ...859...11S} and may appear contradictory to the slope vs. $A_{\text{V}}$ trend, because one expects dust attenuation in the UV to follow the attenuation in the optical. However, as clearly explained in their work, this is only true if the galaxy obeys a universal attenuation law. This is also visible in the plot in Fig. \ref{fig: AV}: for a fixed slope $A_{\text{FUV}}$ indeed increases with $A_{\text{V}}$. When there is a wide range of attenuation slopes, the $A_{\text{FUV}}$ vs. $A_{\text{V}}$ correlation shows a large scatter, as can be seen in Fig. \ref{fig: AFUV_AV}. The ``upper'' data points correspond to the steepest slopes, whereas the ``lower'' points correspond to the shallowest slopes.

The bottom panel of Fig. \ref{fig: AV} shows a subtle trend between the bump strength and $A_{\text{V}}$, in agreement with \citet{2017MNRAS.466.4540H}, \citet{2018ApJ...859...11S} and \citet{2018MNRAS.475.2363T} who also found weaker bumps at higher $A_{\text{V}}$. It is possible that in our case the somewhat larger uncertainty on the derived bump strengths dilutes the correlation. Nevertheless, there is a large variation in bump strengths across the galaxy disc, as can be seen in Fig. \ref{fig: maps}. \cite{2018ApJ...869...70N} showed that the variation in bump strengths can be caused by geometry effects. They found that light scattered into the line-of-sight contributes modestly to reduced bump strengths, but it does not dominate. Moreover, they showed that galaxies with very complex young star-dust geometries (with a larger fraction of unobscured young stars) have reduced bumps. This can indicate that the regions in NGC\,628 with weaker bumps have a significant contribution of unobscured young stars.

\begin{figure}
	\includegraphics[width=\columnwidth]{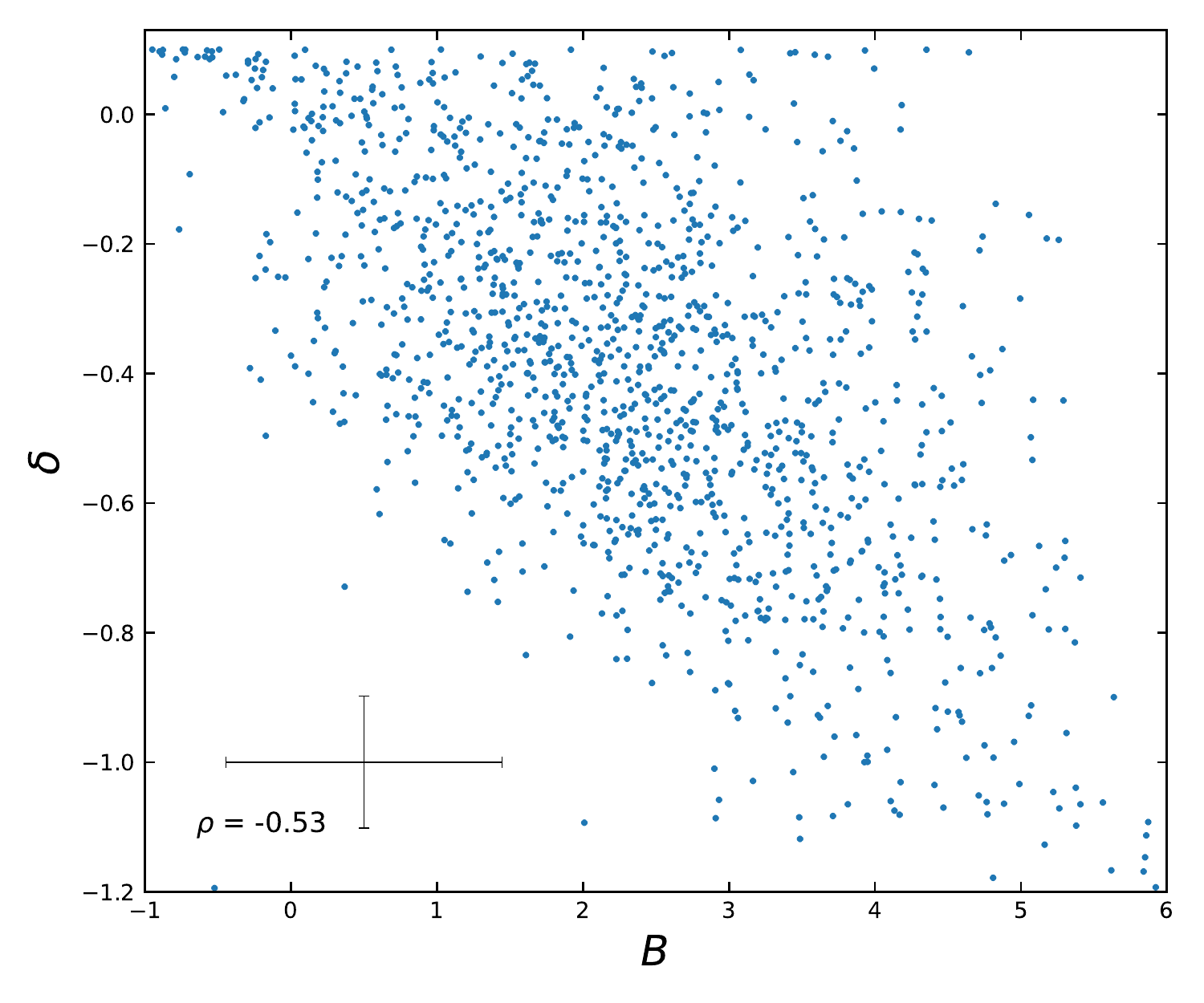}
    \caption{Attenuation slope $\delta$ vs. bump strength $B$. Typical error bars as well as the Spearman correlation coefficient $\rho$ are indicated.}
    \label{fig: bump_slope}
\end{figure}

Plotting the slope versus the bump strength (see Fig. \ref{fig: bump_slope}) results in a modest correlation (although there is a large scatter): steeper curves correspond to stronger bumps. A similar trend was found by \citet{2018ApJ...859...11S}, who showed that the anti-correlation between the bump strength and slope is not driven by modelling artifacts. The same correlation was observed by \cite{2013ApJ...775L..16K} for $0.5<z<2.0$ galaxies and by \cite{2018MNRAS.475.2363T} for $1.5<z<3.0$ star-forming galaxies. However, as stated above, \cite{2018ApJ...869...70N} demonstrated that the fraction of unobscured young stars (and thus the complexity of the star-dust geometry) in their simulations correlates with \textit{both} the slope and the bump strength of the attenuation curve, which leads to a natural relationship between both parameters. They indeed confirm that the slope varies inversely with the bump strength. Finally, \cite{2016ApJ...833..201S} showed that radiative transfer effects lead to shallower attenuation curves with weaker bumps as the interstellar medium is clumpier and dustier.

\begin{figure}
	\includegraphics[width=\columnwidth]{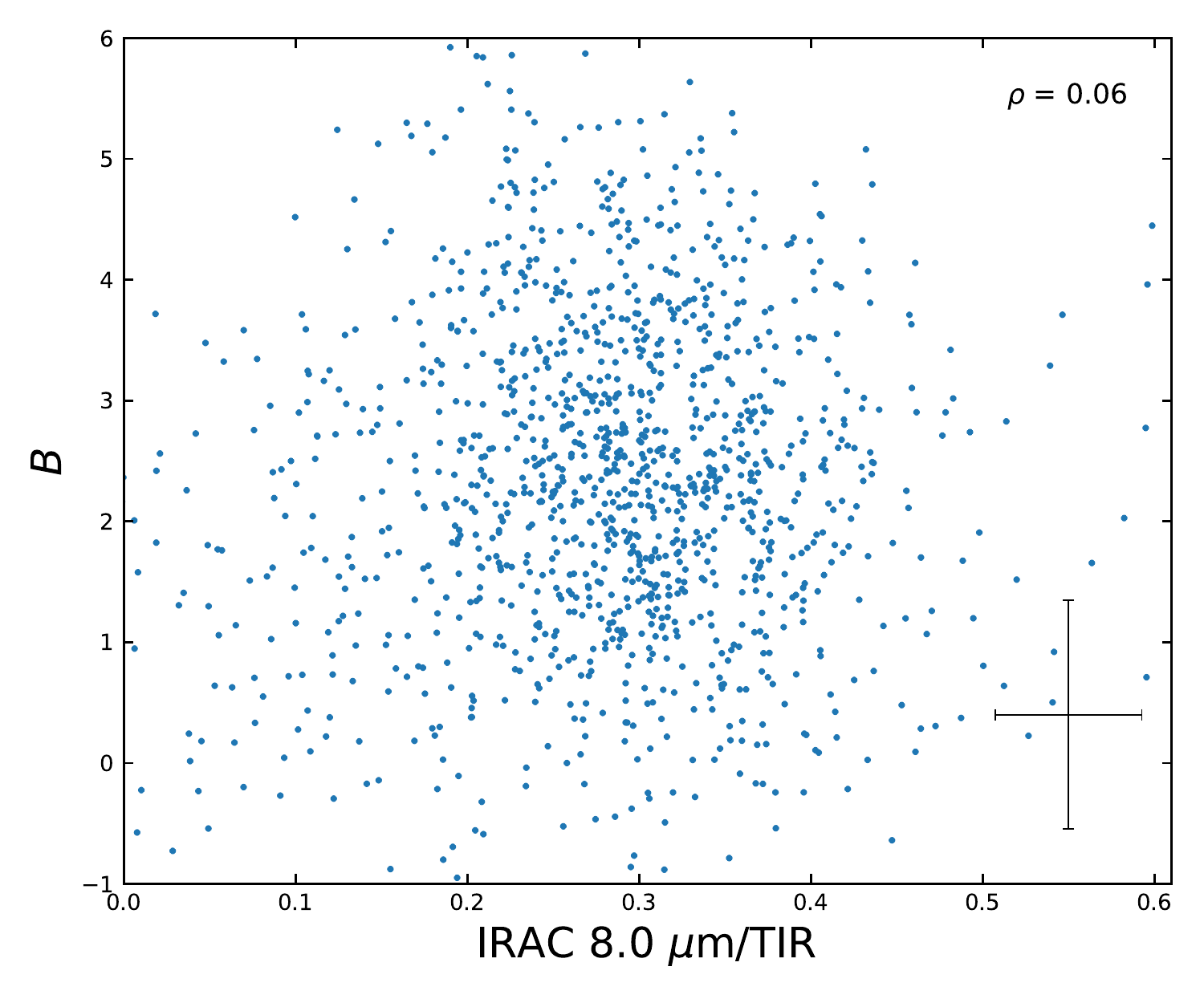}
    \caption{Bump strength $B$ vs. the IRAC\:8\,\micron/TIR fraction. Typical error bars as well as the Spearman correlation coefficient $\rho$ are indicated.}
    \label{fig: bump_IRAC8}
\end{figure}

Several works have pointed to polycyclic aromatic hydrocarbons (PAHs) as bump carriers (e.g. \citealt{1997A&A...323..566L,2007ApJ...657..810D,2008A&A...486L..25C}). If so, the strength of the 2175\,\AA\ bump is expected to correlate with the emission of the mid-infrared PAH features. In Fig. \ref{fig: bump_IRAC8} we show that there is no correlation between the bump strength and the IRAC\:8\,\micron\ emission (corrected for stellar emission) after normalization with the TIR emission. The absence of such a correlation is consistent with the findings from \cite{2012ApJ...760...36P} and \cite{2017MNRAS.466.4540H}. We argue, however, that this result does not necessarily rule out the association of PAHs with the 2175\,\AA\ bump, as (1) the uncertainties on the bump strengths are relatively large, (2) the IRAC\:8\,\micron\ is not a direct measure of the PAH abundance, and can also reflect variations in the radiation field intensity \citep{2010ApJ...724L..44C}, and (3) the bump strength is not a unique measure of the PAH abundance, and is convoluted with the effects of specific mixtures of small and large grains and geometry on the slope.

\subsection{Empirical versus modelling results}
\label{ComparisonColourPlots_vs_Model}
In section \ref{sec: Colour plots} we performed an empirical analysis of the SWIFT and GALEX colours, and compared them to several theoretical stellar spectra attenuated by various dust extinction/attenuation curves. The colours sensitive to the 2175\,\AA\ bump (UVW2-UVM2, UVM2-UVW1) suggest a bump strength in NGC\,628 that resembles that of the MW curve, and strongly rule out the absence of a bump as in the SMC and Calzetti curves. A similar result with a median bump strength of $B=2.31$ (in $A_{\text{V}}>0.2$ regions) (with $B=3$ for the MW) is obtained from more realistic modelling of the SFH and dust attenuation with \textsc{cigale}, although several resolved regions in the galaxy show weaker or stronger bumps. The empirical colour analysis indicates a UV slope steepness in between the MW and SMC dust extinction curves, also in agreement with the median slope value (in $A_{\text{V}}>0.2$ regions) found with \textsc{cigale}: $\delta=-0.37$ (with $\delta \approx -0.45$ for the SMC and $\delta \approx -0.15$ for the MW). However, some regions in the galaxy exhibit a steeper dust attenuation curve than that of the SMC.

As already mentioned in section \ref{sec: Colour plots}, the SSPs used in the empirical colour analysis might not be a good approximation for the resolved stellar populations in NGC\,628. Furthermore, we are attenuating these SSPs with only a few attenuation/extinction curves. When assuming a constant SFR instead of single-age SSPs, the data points in the most constraining colour plots tend to agree more with the median attenuation curve for NGC\,628 obtained with \textsc{cigale}. This indicates that using a continuous SFH is a better approximation for the mix of stellar populations in a resolved galaxy region, and that a more flexible attenuation curve can better reproduce the observations. In addition, by employing the dust energy balance we can break the degeneracies between the age of the stellar populations and the attenuation. For these reasons, we are convinced that the \textsc{cigale} modelling, using a more realistic SFH, and allowing flexibility in the shape of the attenuation curve, gives us more accurate insights into the properties of the attenuation curve on resolved scales in the galaxy, compared to the empirical colour plots.

\subsection{Comparison with similar studies}
\label{sec: comparison}
In this section we consider our results in the context of earlier studies of the dust attenuation in nearby galaxies using SWIFT data. \cite{2011AJ....141..205H} found strong evidence for a MW extinction curve in M81 and Holmberg\,IX. They looked at individual star-forming regions in both galaxies and showed that SMC and Calzetti dust laws are excluded. From the pixel-by-pixel fitting they conclude that both galaxies prefer MW dust across their entire discs with only a small fraction of pixels for which LMC dust is preferred.

\cite{2014MNRAS.440..150H} obtained for the starburst galaxy M82 a strong rejection of extinction curves traditionally used for starburst galaxies \citep{2001NewAR..45..601C}. Overall, the standard MW extinction law is favoured. However, in their revised and more detailed study of M82 \citep{2015MNRAS.452.1412H}, they observed a significant gradient with projected galactocentric distance, with the central starburst region showing a standard MW extinction law with a prominent bump, progressing outwards towards an extinction curve with a weaker bump and steeper slope, tending towards the SMC law. For NGC\,628, on the contrary, we don't see any significant trend between bump or slope and galactocentric distance (apart from the very steep slopes in the galactic centre which might be caused by modelling artifacts). We argue that the lack of any trends with radial distance is because NGC\,628 is a star-forming spiral galaxy with most star formation occurring in the spiral arms, in contrast to M82 which has a strong star-bursting core.

\cite{2017MNRAS.466.4540H} modelled the SMC and discovered that the bump strength has a large-scale gradient across the galaxy, from weaker in the south-west to stronger in the north-east. They also confirmed that the dust extinction curve of the SMC is fairly steep.

\cite{2014ApJ...785..136D} presented a study of the extinction curve within the central 200\,pc region of the M31 bulge. They showed that the extinction curve is generally steep in the circumnuclear region, where the metallicity is super-solar. The derived $R_{\text{V}}=2.4-2.5$ is similar to the one found toward the Galactic bulge. They conclude that large dust grains are destroyed in the harsh environments of the bulge, e.g. via potential shocks from supernova explosions and/or past activities of M31, resulting in steeper curves. Although we argue that the very steep slopes in the centre of NGC\,628 are possibly caused by modelling artifacts, the destruction of large grains in the bulge might also play a role. They additionally derived the extinction curves of five dusty clumps within the central region and encountered significant variations in the mid-UV: some of the extinction curves can be explained by the extinction curve model of \cite{1999PASP..111...63F}, others show an unusually strong 2175\,\AA\ bump, which is weak elsewhere in the M31 disc \citep{1996ApJ...471..203B}.

Although the sample of nearby galaxies studied on resolved scales is currently limited, the observed range of dust attenuation curves (with variable ranges of bump strengths and UV slopes) for this small set of galaxies allows us to conclude that the shape of the dust law varies from one galaxy to another, and also shows variations on resolved galactic scales. For NGC\,628, we show that geometry effects possibly dominate the variations in the slopes and bump strengths across the disc of the galaxy. If this trend persists, we speculate that variations in the shape of dust attenuation curves in nearby galaxies could be driven mostly by geometrical effects. We caution, however, that the modelling techniques used in these various works make different assumptions about the SFH, the dust attenuation curves and the stellar populations, which might impact a direct comparison of the results. A consistent study of the dust attenuation curve in a statistical sample of nearby galaxies, for which the modelling has been done in a homogeneous way within the DustKING framework, will allow us to disentangle whether variations in dust attenuation laws are driven by intrinsic grain properties, geometry effects and/or different modelling techniques (Decleir et al, in prep.). 
 
\section{Summary and conclusions}
\label{sec: Summary}
In this work, we studied the dust attenuation properties on resolved scales of about 325\,pc in the nearby spiral galaxy NGC\,628 using a FUV-to-FIR multi-wavelength dataset including SWIFT UVOT observations that probe the 2175\,\AA\ dust absorption feature. We presented a new data reduction pipeline for the SWIFT UVOT images, which is fully optimized for extended sources, and accounts for the coincidence loss affecting the photon-counting instrument on a pixel-by-pixel basis.

We have utilised the SED fitting code \textsc{cigale} to constrain the SFH and the shape of the dust attenuation curve (i.e. UV slope, bump strength and dust attenuation) based on a set of photometric measurements in the FUV-to-NIR wavelength domain. In addition, we have employed the total-infrared luminosity as an extra constraint in \textsc{cigale} to limit degeneracies between the age of stellar populations and dust attenuation parameters. With this novel SED modelling technique, we were able to constrain the attenuation curve slope and bump strength reasonably well (with median relative uncertainties of 27\% and 41\% on the slopes and bump strengths, respectively, on resolved scales).

\noindent The main conclusions of our SED modelling with \textsc{cigale} are presented here:
\begin{itemize}
\item The median attenuation curve of NGC\,628 has a fairly steep slope ($\delta = -0.37 \pm 0.10$) in between that of the MW ($\delta \approx -0.15$) and the SMC ($\delta \approx -0.45$) curves and a bump ($B = 2.31 \pm 0.95 $) somewhat smaller compared to the MW ($B=3$). 
\item The shape of the global dust attenuation law of NGC\,628 (with $B=2.24\pm1.16$ and $\delta=-0.55\pm0.15$) is not very different from the shape of the median curve from the pixel-by-pixel fitting. Thus, for this galaxy, an investigation of the attenuation curve on resolved scales gives similar results to our finding on a global galaxy scale.
\item Shallower slopes are observed in regions with higher levels of dust attenuation, reflecting the dominance of absorption over scattering events, resulting in a flattening of the dust law \citep{2013MNRAS.432.2061C}. Geometry effects seem to dominate the variations in the slope.
\item The weak observed trend of smaller bumps in regions with more dust attenuation (higher $A_V$) can also be attributed to geometry effects \citep{2018ApJ...869...70N}.
\item There is a trend between the slope and the bump strength of the attenuation curve: steeper curves exhibit stronger bumps. Geometrical effects can lead to a natural relationship between both parameters \citep{2018ApJ...869...70N}.
\item No trend between the bump strength and the IRAC\:8\,\micron/TIR ratio could be found. We argue that this does not necessarily rule out PAHs as bump carriers, if the IRAC\:8\,\micron\ emission is not a good proxy of the PAH abundance and/or if the bump strength is not an accurate measure of the absolute dust absorption at 2175\,\AA. In addition, the uncertainties on the bump strengths are substantial.
\end{itemize}
 
The median dust attenuation parameters obtained with \textsc{cigale} are mostly in line with the conclusions inferred from the empirical colour analysis (see section \ref{sec: Colour plots}), although with \textsc{cigale} we find a larger variety in attenuation slopes and bump strengths across the galaxy disc. We argue that the use of more complex SED models, which are able to account for a range of stellar populations (through the SFH), and a flexible parameterization of the dust law gives more accurate insights into the dust attenuation properties.

In comparison with previous resolved studies of the dust attenuation law in M81, M82 and the SMC and LMC, the bump strength in NGC\,628 is consistent with the MW-type bump in M81, M82 and some regions of the SMC/LMC. The UV slope is significantly steeper than the slopes inferred for M81 and M82, but not as steep as the average SMC curve. To understand how the dust attenuation curve varies in the nearby Universe, and how model assumptions affect the inference of dust parameters, we require a study of the dust attenuation curve in a statistical sample of nearby galaxies, for which the modelling has been done in a homogeneous way. In future work, we will extend the analysis presented here for NGC\,628 to a larger sample of nearby galaxies (Decleir et al., in prep.), which will allow us to disentangle variations in the dust attenuation law from potential model biases.

\section*{Acknowledgements}
The authors would like to express a special thanks to Michael Brown and Alice Breeveld for their help with the calibration of the SWIFT images. We would also like to thank Bryan Irby and Robert Wiegand from the SWIFT helpdesk.

This work benefits from a PhD Fellowship of the Research Foundation - Flanders (FWO-Vlaanderen).
I. De Looze gratefully acknowledges the support of the Research Foundation - Flanders. M. Boquien was supported by MINEDUC-UA projects, code ANT 1655 and ANT 1656, and FONDECYT project 1170618.

This research has made use of the SVO Filter Profile Service, supported from the Spanish MINECO through grant AyA2014-55216.

%%%%%%%%%%%%%%%%%%%%%%%%%%%%%%%%%%%%%%%%%%%%%%%%%%


\bibliographystyle{mnras}
\bibliography{References}

%%%%%%%%%%%%%%%%%%%%%%%%%%%%%%%%%%%%%%%%%%%%%%%%%%

%%%%%%%%%%%%%%%%% APPENDICES %%%%%%%%%%%%%%%%%%%%%

\appendix

\section{IDs of SWIFT UVOT images}
\label{app: ID}
Table \ref{tab: SWIFT_ID} lists the IDs of the SWIFT UVOT images that were used in this work (see section \ref{sec: aspect} for more information). Note that each of these ``images'' contains several individual frames that need to be reduced separately.

\begin{table}
	\centering
	\caption{Image IDs of the SWIFT UVOT images that were used in this work.}
	\label{tab: SWIFT_ID}
	\begin{tabular}{lllp{3.5cm}}
		\hline
		epoch & filter & prefix & image IDs\\
		\hline
        2007 & UVW2 & sw000365680 & 01\\
        & UVM2 & sw000365680 & 01\\
        & UVW1 & sw000365680 & 01\\
        \hline
        2008 & UVW2 & sw000358680 & 01, 02\\
        & UVM2 & sw000358680 & 01, 02\\
        & UVW1 & sw000358680 & 01, 02\\
        \hline
        2013 & UVW2 & sw000328910 & 01, 02, 03, 06, 09, 11, 14, 15, 16, 19, 20, 21, 22, 23, 24, 25, 27, 29, 31, 33, 35, 37, 39, 40, 42, 43, 45, 46, 47\\
		& UVM2 & sw000328910 & 01, 02, 12, 17, 18, 19, 21, 22, 23, 25, 27, 29, 31, 33, 35, 37, 40, 42, 43, 45, 46, 47\\
        & UVW1 & sw000328910 & 03, 10, 19, 20, 21, 22, 23, 24, 25, 27, 29, 31, 33, 35, 37, 40, 42, 45\\
        \hline
        2015 & UVW2 & sw000365680 & 03\\
		& UVW1 & sw000365680 & 02\\
		\hline
	\end{tabular}
\end{table}

\section{Calculation of the coincidence loss correction}
\label{app: coicorr}

Since the existing calculations of coincidence loss corrections in SWIFT UVOT images are not feasible for pixel-by-pixel studies of extended sources, as described in section \ref{subsec: coiloss}, we had to come up with a new strategy. We adjusted the correction technique for point sources of \cite{2008MNRAS.383..627P} to a pixel-by-pixel algorithm. Because of the photon splash that is created in the detector, one cannot simply calculate a correction factor for each individual pixel without taking into account the surrounding pixels. Therefore, we determine the coincidence loss correction factor within a $9\times9$ pixels sized box (or $81\arcsec^2$ in our case) centered on the pixel of interest, which covers an area equivalent to a $R=5\arcsec$ aperture region. The coincidence-loss-corrected count rate $C_{\textrm{coicorr}}$ (in counts/s) in a certain pixel is calculated as
\begin{equation}
C_{\textrm{coicorr}} = C_{\textrm{obs}} \times f_{\textrm{coicorr}}
\end{equation}

\noindent where $C_{\textrm{obs}}$ is the observed count rate in that pixel and $f_{\textrm{coicorr}}$ is the coincidence loss correction factor determined within a $9\times9$ pixels sized box centered on that pixel. Following the prescriptions by \cite{2008MNRAS.383..627P}, the correction can be calculated as follows.

The theoretical coincidence-loss-corrected count rate $C_{\textrm{theory}\square}$\footnote{The $\square$ indicates that we are working in a $9\times9$ pixels sized box.} (in counts/s) in a $9\times9$ pixels box is:

\begin{equation}
C_{\textrm{theory}\square} = \frac{- \ln (1 - \alpha \:C_{\textrm{raw}\square} f_t)}{\alpha f_t}
\end{equation}

\noindent where $C_{\textrm{raw}\square}$ is the raw observed  count rate in the $9\times9$ pixels box. $f_t$ is the frame time (0.01103\,s for a full frame; keyword FRAMTIME in the FITS file header) and $\alpha$ is the dead time correction factor (1 minus the dead time fraction; 0.9842 for a full frame; keyword DEADC in the FITS file header). As already explained, this theoretical coincidence loss expression cannot be applied on a pixel-by-pixel basis because every photon count is assigned to a certain pixel by centroiding a photon splash that was spread over five physical CCD pixels. To compensate for this, we need to use the empirical polynomial correction from \cite{2008MNRAS.383..627P} to account for the differences between the observed and theoretical coincidence loss correction:
\begin{equation}
\label{eq: f(X)}
f (x) = 1 + a_1 x + a_2 x^2 + a_3 x^3 + a_4 x^4
\end{equation}
where $x = C_{\textrm{raw}\square} f_t$. The coefficients in this equation are taken from Table 1 of the SWIFT UVOT CALDB Release Note 14-R02\footnote{\url{https://heasarc.gsfc.nasa.gov/docs/heasarc/caldb/swift/docs/uvot/uvot_caldb_coi_02.pdf}}. The full coincidence-loss-corrected incident count rate in the $9\times9$ pixels box $C_{\textrm{corr}\square}$ (in counts/s) is then:
\begin{equation}
C_{\textrm{corr}\square} = C_{\textrm{theory}\square} \times f (x)
\end{equation}

The coincidence loss correction factor $f_{\textrm{coicorr}}$ in the central pixel of the box  is then eventually:
 \begin{equation}
f_{\textrm{coicorr}} = \frac{C_{\textrm{corr}\square}}{C_{\textrm{raw}\square}}
\end{equation}

We refer the reader to section 7 of \cite{2008MNRAS.383..627P} for more details on the calculation.

To get an estimate of the uncertainty on the coincidence loss correction, we determine the standard deviation on the count rate $\sigma_\square$ in each $9\times9$ pixels box. Following the same procedure as above, we calculate a ``minimum'' correction factor, based on a minimum count rate in the box $C_{\textrm{min}\square}=C_{\textrm{raw}\square}-81\times\sigma_\square$ and a ``maximum'' correction factor, based on a maximum count rate $C_{\textrm{max}\square}=C_{\textrm{raw}\square}+81\times\sigma_{\square}$. Subsequently, we compute the difference between the coincidence-loss-corrected flux and the minimum/maximum corrected flux in every pixel. The largest of these two differences can be considered as an upper limit of the uncertainty on the corrected flux. In our case, typical uncertainties are extremely small (<0.2\%) compared to other sources of uncertainty and can therefore safely be neglected.

\section{Parameter ranges used in the \textsc{cigale} fitting}
\label{app: cigale_par}
Table \ref{tab: cig_params} lists the parameters that were used in the \textsc{cigale} fitting, with a range of values in the case of free parameters or with one value in the case of fixed parameters.

\begin{table*}
	\centering
	\caption{Parameter ranges used in the \textsc{cigale} fitting for the SFH: the e-folding time of the main stellar population model $\tau_{\textrm{main}}$ (in Myr), the age of the oldest stars in the galaxy $a$ (in Myr), the age of the drop in the star formation activity $a_{\textrm{trunc}}$ (in Myr), the ratio between the SFR after and before truncation $r_{\textrm{sfr}}$, the multiplicative factor controlling the amplitude of SFR $A_{\textrm{sfr}}$, and the normalization flag $norm$ to normalize the SFH to produce one solar mass; and the dust attenuation: the amplitude of the UV bump $B$, the powerlaw slope $\delta$, the colour excess of the stellar continuum light $E(B-V)$, the reduction factor for the colour excess of the old population compared to the young one $E(B-V)_{\textrm{old factor}}$, the central wavelength of the UV bump $\lambda$ (in nm), and the width of the UV bump $FHWM$ (in nm). All other (fixed) parameter values are also given: the initial mass function $IMF$, the metallicity $Z$, the age separation between the young and old stellar populations $a_{\textrm{sep}}$ (in Myr), the ionization parameter of the nebular emission $log(U)$, the fraction of Lyman continuum photons escaping the galaxy $f_{\textrm{esc}}$, the fraction of Lyman continuum photons absorbed by dust $f_{\textrm{dust}}$, the line width $lw$ (in km/s), the flag $em$ whether or not to include nebular emission, the AGN fraction $f_{\textrm{AGN}}$, the slope of the \textsc{dale2014} dust emission templates $\alpha$, and the redshift $z$.}
	\label{tab: cig_params}
	\begin{tabular}{lp{14.5cm}}
		\hline
        \multicolumn{2}{c}{\textit{SFH parameters}}\\
        \hline
        $\tau_{\textrm{main}}$ & 500, 1000, 2000, 4000, 5000, 6000, 7000, 8000\\
        $a$ & 8000, 9000, 10000, 11000, 12000\\
        $a_{\textrm{trunc}}$ & 100\\
        $r_{\textrm{sfr}}$ & 0, 0.001, 0.01, 0.02, 0.05, 0.075, 0.1, 0.2, 0.5, 0.75, 1, 2, 5, 7.5, 10\\
        $A_{\textrm{sfr}}$ & 1\\
        $norm$ & True\\
        \hline
		\multicolumn{2}{c}{\textit{dust attenuation parameters}}\\
        \hline
        $B$ & -1, -0.5, 0, 0.5, 1, 1.5, 2, 2.5, 3, 3.5, 4, 4.5, 5, 5.5, 6\\
        $\delta$ & -1.2, -1.15, -1.1, -1.05, -1, -0.95, -0.9, -0.85, -0.8, -0.75, -0.7, -0.65, -0.6, -0.55, -0.5, -0.45, -0.4, -0.35, -0.3, -0.25, -0.2, -0.15, -0.1, -0.05, 0, 0.05, 0.1\\
        $E(B-V)$ & 0, 0.01, 0.02, 0.03, 0.04, 0.05, 0.06, 0.07, 0.08, 0.09, 0.1, 0.125, 0.15, 0.175, 0.2, 0.225, 0.25, 0.275, 0.3, 0.325, 0.35, 0.375, 0.4, 0.425, 0.45, 0.475, 0.5\\
        $E(B-V)_{\textrm{old factor}}$ & 1\\
        $\lambda$ & 217.5\\
        $FWHM$ & 35\\
		\hline
        \multicolumn{2}{c}{\textit{other (fixed) parameters}}\\
		\hline
        $IMF$ & 1 (Chabrier)\\
        $Z$ & 0.02\\
        $a_{\textrm{sep}}$ & 10\\
        $log(U)$ & -3.0\\
        $f_{\textrm{esc}}$ & 0\\
        $f_{\textrm{dust}}$ & 0\\
        $lw$ & 300\\
        $em$ & True\\
        $f_{\textrm{AGN}}$ & 0\\
        $\alpha$ & 2\\
        $z$ & 0\\
  	\end{tabular}
\end{table*}

\section{Observations, models and residuals for all wavebands fitted with \textsc{cigale}}

Fig. \ref{fig: UV_maps} shows the observations, the models and the residuals for the 15 wavebands that were fitted with \textsc{cigale}. Residuals are calculated as the relative difference between the modelled and the observed flux. In the histograms of the residuals we draw vertical lines at typical uncertainties on the observed flux, e.g. around 7\% for the FUV band. Overall, the observations are fitted quite well with residuals within the typical uncertainties on the observed flux. In the SWIFT UVW2 and UVW1 bands the model slightly underestimates the observations (negative residuals), with the largest offsets (around 15\%) in the inter-arm regions where the uncertainties on the flux are larger (lower SNR). The SDSS z-band shows a somewhat larger offset compared to the other optical bands, which might be due to issues with the background subtraction in this image. Finally, the 2MASS bands are fitted quite well in the central regions of the galaxy, but worse in the outer regions due to a lower SNR. Typical uncertainties on the observed fluxes in these bands are also very high (36-90\% on average).

\label{app: figures}
\begin{figure*}
	\includegraphics[height=22.7cm]{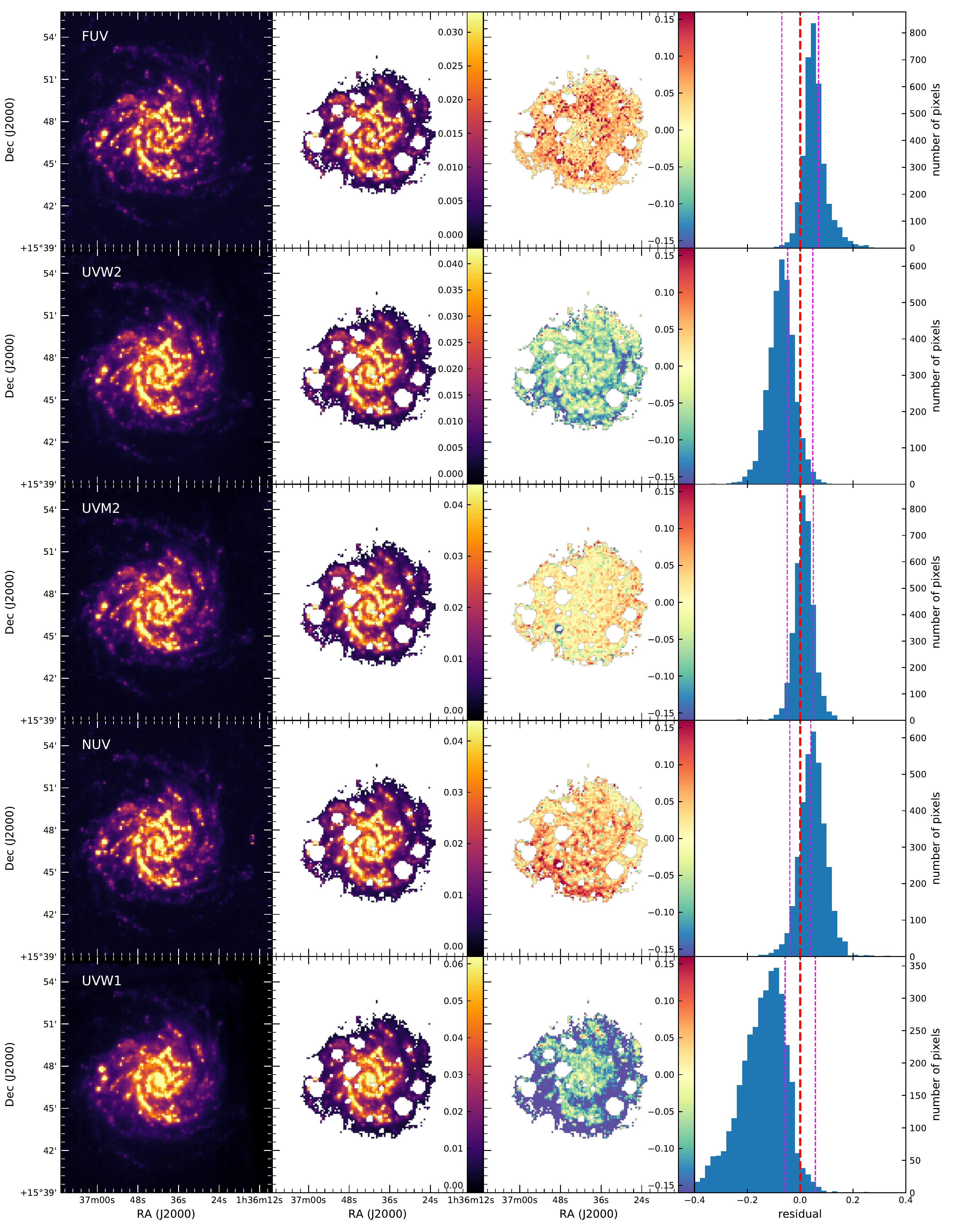}
    \caption{From left to right: Observation and model maps (in MJy/sr), residual map and histogram of the residual for the UV bands fitted with \textsc{cigale}. White regions within the galaxy models and residuals are masked pixels of foreground stars that were left out from the fitting. The dashed magenta lines in the histograms represent the median uncertainty on the observed flux.}
    \label{fig: UV_maps}
\end{figure*}

\begin{figure*}
	\ContinuedFloat
	\includegraphics[height=23cm]{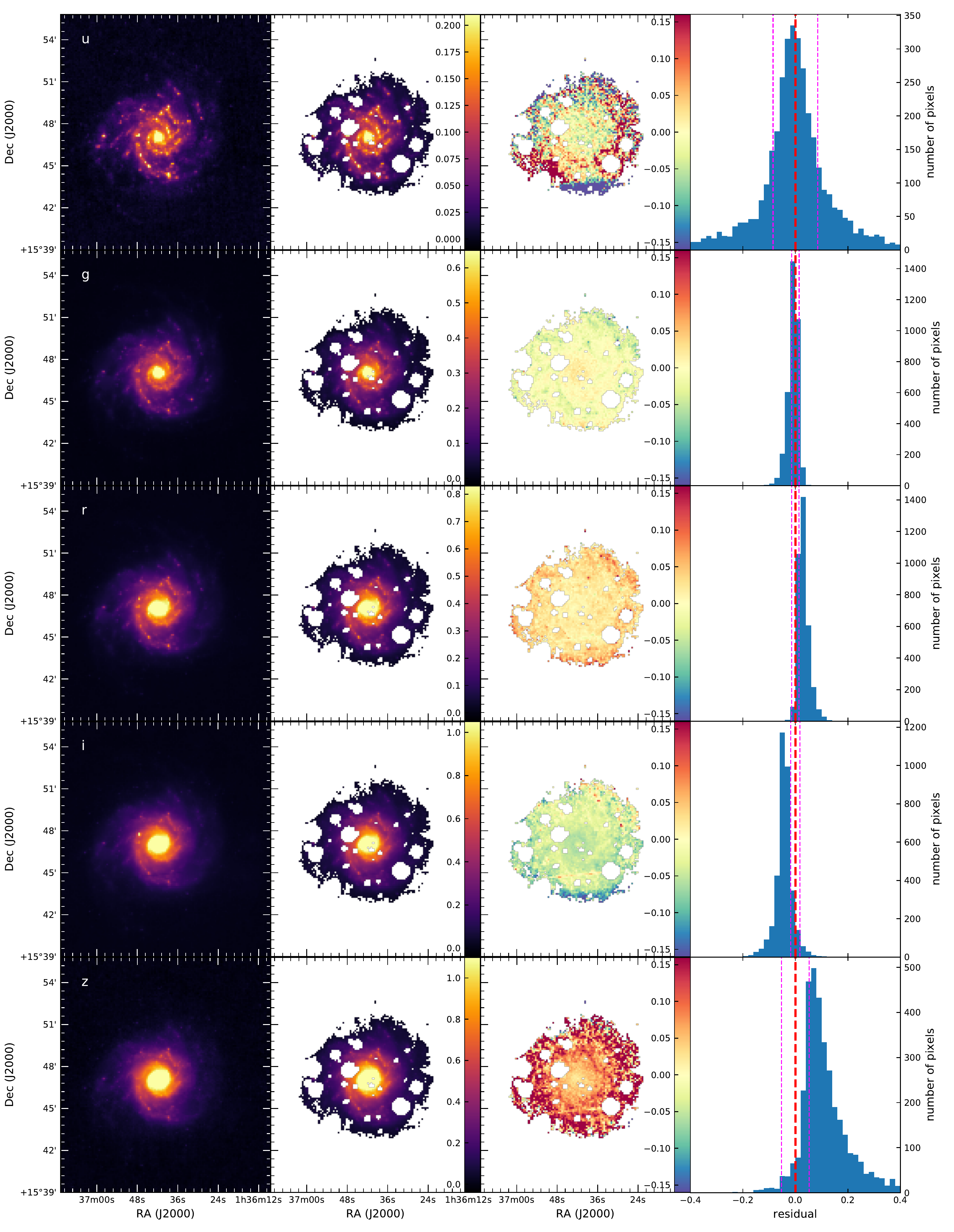}
    \caption{Continued: Optical bands.}
    \label{fig: opt_maps}
\end{figure*}

\begin{figure*}
	\ContinuedFloat
	\includegraphics[height=23cm]{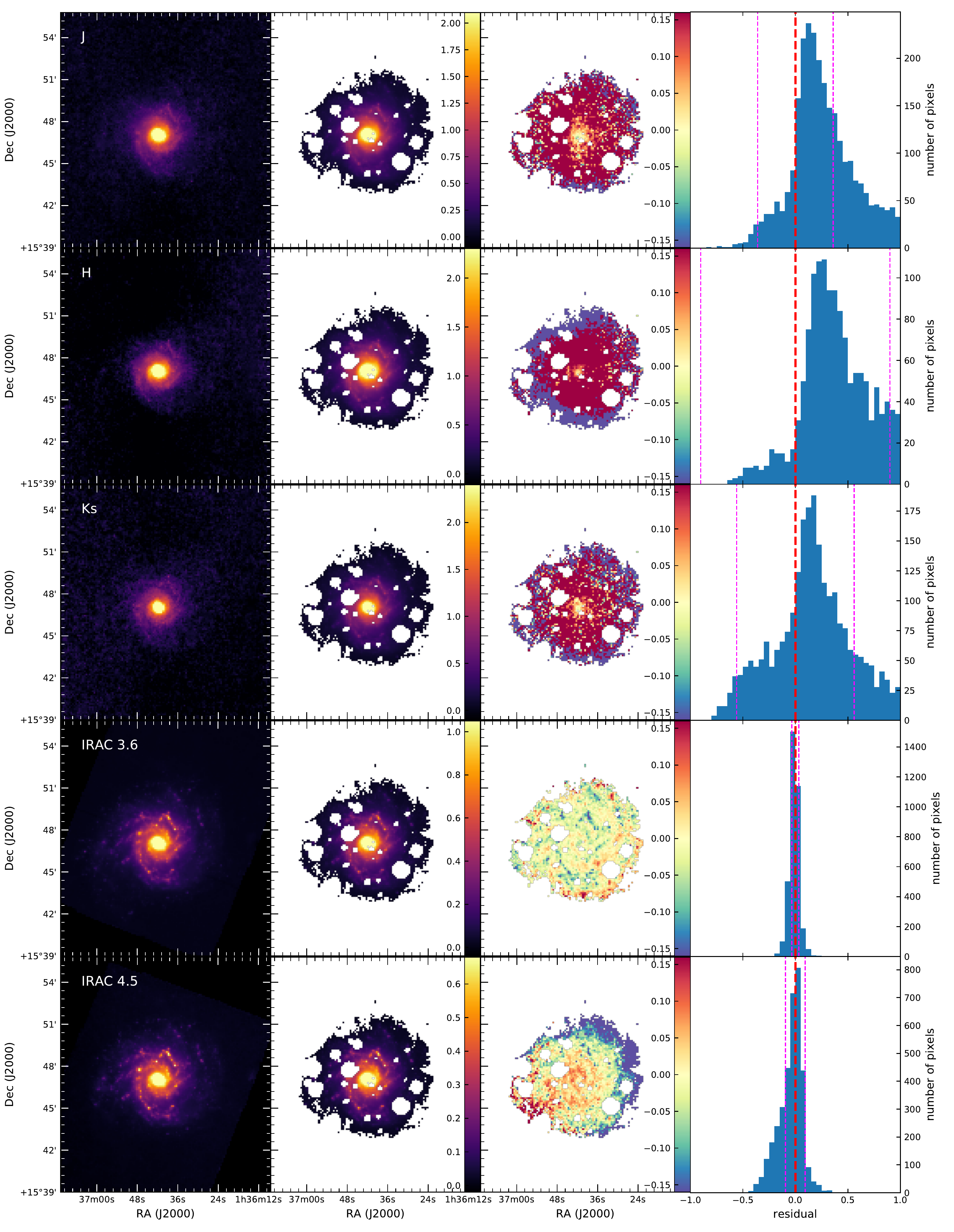}
    \caption{Continued: NIR bands.}
    \label{fig: NIR_maps}
\end{figure*}

\section{Mock testing with \textsc{cigale}}
\label{app: mock}
\begin{figure*}
	\includegraphics[height=14cm]{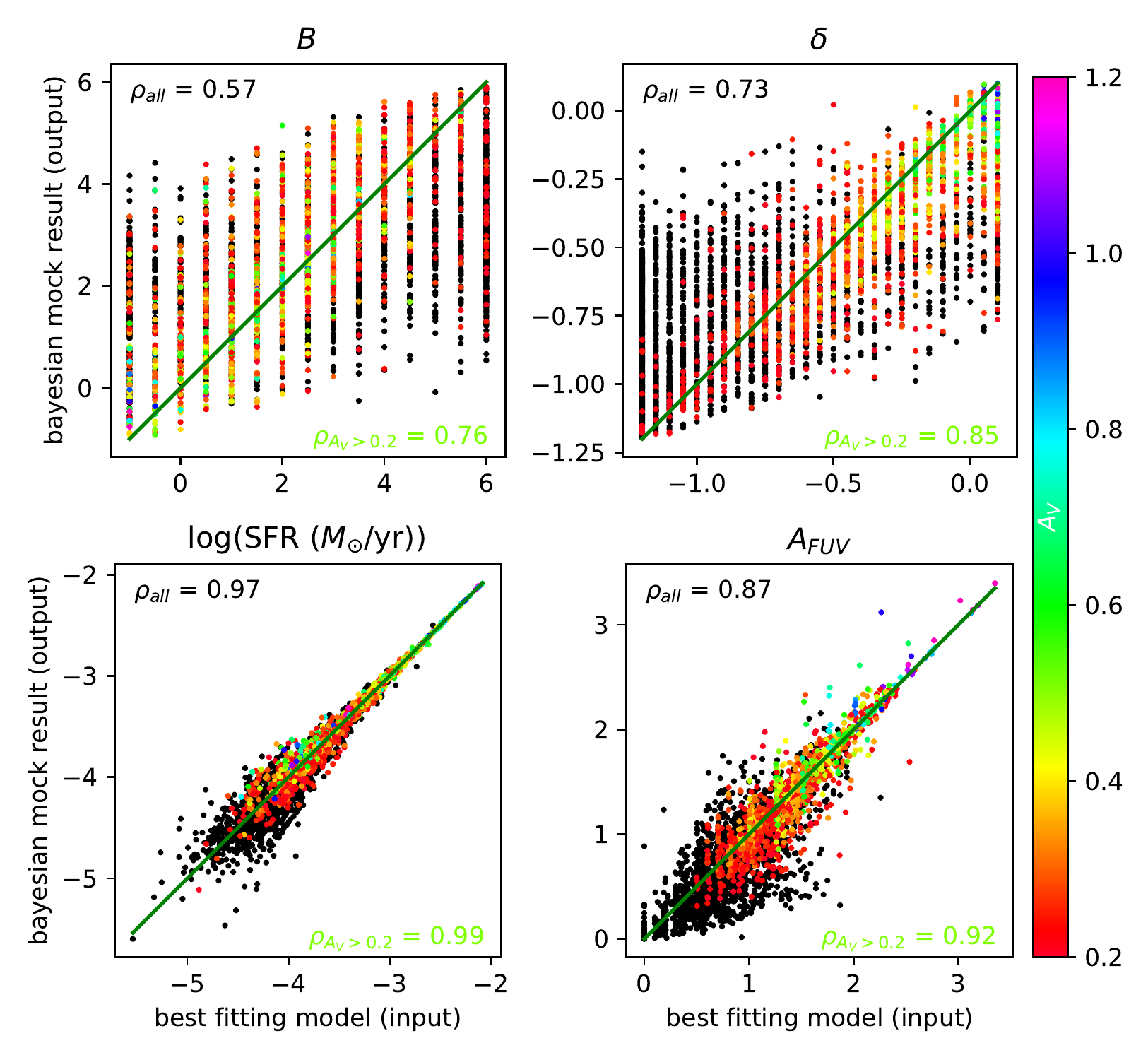}
    \caption{Mock data analysis results for the UV bump strength $B$, the attenuation slope $\delta$, the SFR (in $M_{\odot}$/yr) and the FUV attenuation $A_{\text{FUV}}$. The black dots represent pixels with $A_{\text{V}}\leq0.2$. The Spearman correlation coefficient $\rho$ is shown for all pixels (in black) and for pixels with $A_{\text{V}}>0.2$ (in green).}
    \label{fig: mock}
\end{figure*}

In order to verify whether our \textsc{cigale} model is able to constrain physical properties such as the SFR and the shape of the dust attenuation curve, we performed a mock data analysis with input data generated from model SEDs for which we know the true parameters and by assuming the same noise levels that affect our observations in each waveband. To this aim, we followed a similar procedure as in appendix B.2 of \cite{2016A&A...591A...6B}. We start from the SED and the associated physical properties of the best-fitting model obtained for NGC\,628. A random Gaussian noise is added to the flux in each band, based on the 1$\sigma$ uncertainty on the observed fluxes, to simulate an observation of the SED. In a second step, this simulated observation is fitted with the bayesian approach of \textsc{cigale}. In Fig. \ref{fig: mock} we compare the ``true'' (input) and ``estimated'' (output) values for the UV bump strength $B$, the attenuation slope $\delta$, the SFR and the FUV attenuation $A_{\text{FUV}}$. 

It is clear that the SFR and the $A_{\text{FUV}}$ are constrained well; all mock models are concentrated along the one-to-one correlation with a scatter of only 0.12 on the log(SFR($M_{\odot}$/yr)) and 0.25 on the $A_{\text{FUV}}$. On the other hand, the bump strength and UV slope seem to be more difficult to reproduce. However, ignoring pixels with an $A_{\text{V}}\leq0.2$ (indicated in black in Fig. \ref{fig: mock}) significantly improves the constraints on these parameters, with the scatter around the one-to-one relation on the bump strength decreasing from 1.88 to 1.23, and on the slope from 0.28 to 0.18. For those pixels with a very low dust attenuation it is essentially impossible to get meaningful constraints on the attenuation curve. In addition, we applied a correction to the images to account for the Galactic foreground extinction of $A_{\text{V}}=0.188$ (see section \ref{sec: Data processing}). This will add noise to the data points making them less reliable, especially for pixels with $A_{\text{V}}\leq0.2$. The results improve further with increasing $A_{\text{V}}$: the scatter on the bump, e.g., decreases to 0.90 if we only consider mock data points with $A_{\text{V}}>0.5$, and on the slope to 0.09. Due to the non-negligible uncertainties on the resolved SWIFT fluxes (4.5-6\% on average), it is not surprising that the bump strength is only well constrained for pixels with sufficiently high dust attenuation. Therefore, we are confident that the model is able to constrain all properties within reasonable uncertainties.

\begin{figure*}
	\includegraphics[height=14cm]{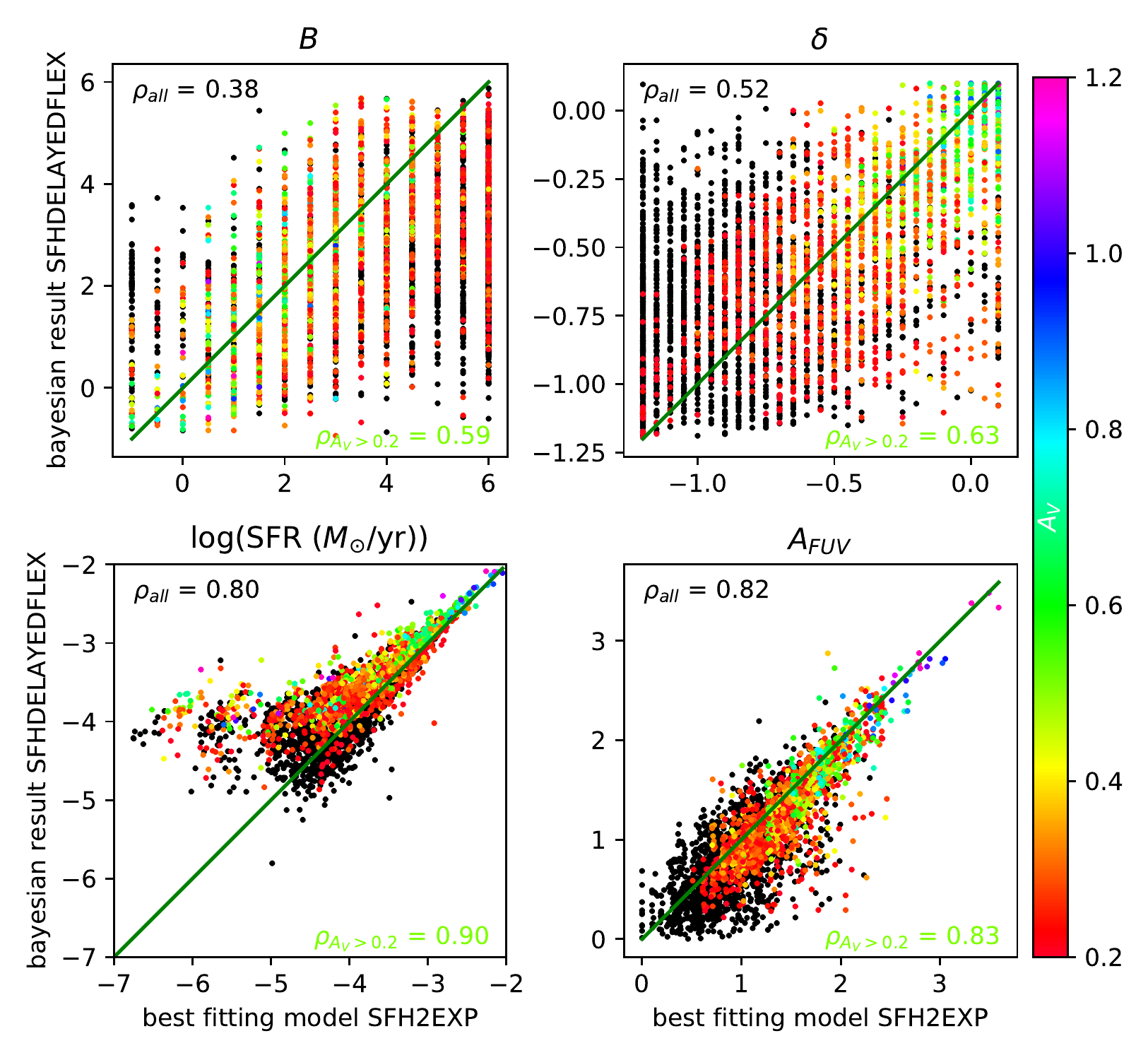}
    \caption{Mock data analysis results for the UV bump strength $B$, the attenuation slope $\delta$, the SFR (in $M_{\odot}$/yr) and the FUV attenuation $A_{\text{FUV}}$, for the hybrid approach as explained in the text. The black dots represent pixels with $A_{\text{V}}\leq0.2$. The Spearman correlation coefficient $\rho$ is shown for all pixels (in black) and for pixels with $A_{\text{V}}>0.2$ (in green).}
    \label{fig: hybrid}
\end{figure*}

In the first mock test (see Fig. \ref{fig: mock}), we have used the same set-up for the SFH for both the generation and the fitting of the mock data in \textsc{cigale}, which does not allow to test the influence of our assumed SFH. To check the dependence of the results on the model SFH (in addition to the test explained in section \ref{sec: test}) we also performed a ``hybrid'' mock data analysis in which we start from the best model for NGC\,628 obtained with a double exponential SFH (\textsc{sfh2exp}) to create a mock dataset, which is subsequently fitted using the delayed and flexible SFH (\textsc{sfhdelayedflex}). The comparison between the ``true'' and ``estimated'' properties is given in Fig. \ref{fig: hybrid}. The $A_{\text{FUV}}$ is again constrained very well with a scatter of 0.31 around the one-to-one correlation, while there is somewhat more scatter (0.60) on the log(SFR($M_{\odot}$/yr)). The bump and the slope show substantial scatter (2.32 and 0.35, respectively), which improves down to a dispersion of 1.02 on the bump and 0.15 on the slope considering only models with $A_{\text{V}}>0.5$. The larger scatter on the SFR indicates that our assumption on the SFH will affect the modelling of the unattenuated stellar population, which will in turn alter the derivation of the dust attenuation curve parameters. For pixels with $A_{\text{V}}>0.5$, the SFR and UV slope are reproduced relatively well, whereas the bump strength is more difficult to constrain. This hybrid mock analysis demonstrates that the assumed SFH has an influence on the results. However, as explained in section \ref{sec: test}, we argue that the delayed and flexible SFH is more adequate to model the resolved regions in NGC\,628 compared to the double exponentially declining SFH. The delayed SFH with a final burst/quench is better adapted because it can more easily represent the more diverse SFHs we expect at a local scale in the galaxy disc: spiral arms had a recent episode of star formation, whereas star formation in interarm regions has fallen. This cannot be obtained with a double exponentially declining SFH.

%%%%%%%%%%%%%%%%%%%%%%%%%%%%%%%%%%%%%%%%%%%%%%%%%%

% Don't change these lines
\bsp	% typesetting comment
\label{lastpage}
\end{document}